\documentclass[conference]{IEEEtran}
\pdfoutput=1
\IEEEoverridecommandlockouts
\usepackage{cite}
\usepackage{amsmath,amssymb,amsfonts}
\usepackage{bm}
\usepackage{booktabs}
\usepackage{dcolumn}
\usepackage{verbatim}
\usepackage{siunitx}
\usepackage{algpseudocode}
\usepackage{listings}
\usepackage{graphicx}
\usepackage{textcomp}
\usepackage{xcolor}
\def\BibTeX{{\rm B\kern-.05em{\sc i\kern-.025em b}\kern-.08em
    T\kern-.1667em\lower.7ex\hbox{E}\kern-.125emX}}
\usepackage{soul}
\newcommand{\revised}[2]{#1}
\usepackage[hidelinks,pdfa]{hyperref}
\usepackage{orcidlink}
\usepackage{hyperxmp}
\hypersetup{%
  pdfapart=3,
  pdfaconformance=B,
  pdflicenseurl={http://creativecommons.org/licenses/by/4.0/},
  pdftitle={Performance-portable solid mechanics via matrix-free p-multigrid},
  pdfauthor={Jed Brown, Valeria Barra, Natalie Beams, Leila Ghaffari, Matthew Knepley, William Moses, Rezgar Shakeri, Karen Stengel, Jeremy L. Thompson, Junchao Zhang}
}

\def \diff {\operatorname{d}\!}
\def \tcolon {\!:\!}
\def \trace {\operatorname{trace}}
\newcommand{\code}[1]{\textsf{#1}}

\newcommand{\DoF}[1]{\qty{#1}{DoF}}

\newcommand{\MDoF}[1]{\qty{#1}{MDoF}}

\newcommand{\GDoF}[1]{\qty{#1}{GDoF}}
\newcommand{\GDoFs}[1]{\qty{#1}{GDoF\per\second}}

\usepackage{fancyhdr}
\begin{document}

\title{
  Performance-Portable Solid Mechanics via Matrix-Free $p$-Multigrid
}


\author{
\IEEEauthorblockN{Jed Brown\,\orcidlink{0000-0002-9945-0639}}
\IEEEauthorblockA{\textit{University of Colorado, Boulder} \\
jed@jedbrown.org}
\and
\IEEEauthorblockN{Valeria Barra\,\orcidlink{0000-0003-1129-2056}}
\IEEEauthorblockA{\textit{California Institute of Technology} \\
valeria@caltech.edu}
\and
\IEEEauthorblockN{Natalie Beams\,\orcidlink{0000-0001-6060-4082}}
\IEEEauthorblockA{\textit{University of Tennessee} \\
nbeams@icl.utk.edu}
\and
\IEEEauthorblockN{Leila Ghaffari\,\orcidlink{0000-0002-0965-214X}}
\IEEEauthorblockA{\textit{University of Colorado, Boulder} \\
leila.ghaffari@colorado.edu}
\and
\IEEEauthorblockN{Matthew Knepley\,\orcidlink{0000-0002-2292-0735}}
\IEEEauthorblockA{\textit{University of Buffalo} \\
knepley@gmail.com}
\and
\IEEEauthorblockN{William Moses\,\orcidlink{0000-0003-2627-0642}}
\IEEEauthorblockA{\textit{Massachusetts Institute of Technology} \\
wmoses@mit.edu}
\and
\IEEEauthorblockN{Rezgar Shakeri\,\orcidlink{0000-0003-4790-7016}}
\IEEEauthorblockA{\textit{University of Colorado, Boulder} \\
rezgar.shakeri@colorado.edu}
\and
\IEEEauthorblockN{Karen Stengel\,\orcidlink{0000-0002-8764-9122}}
\IEEEauthorblockA{\textit{University of Colorado, Boulder} \\
karen.stengel@colorado.edu}
\and
\IEEEauthorblockN{Jeremy L. Thompson\,\orcidlink{0000-0003-2980-0899}}
\IEEEauthorblockA{\textit{University of Colorado, Boulder} \\
jeremy.thompson@colorado.edu}
\and
\IEEEauthorblockN{Junchao Zhang\,\orcidlink{0000-0003-0367-2358}}
\IEEEauthorblockA{\textit{Argonne National Laboratory} \\
jczhang@mcs.anl.gov}
}

\maketitle

\begin{abstract}
  Finite element analysis of solid mechanics is a foundational tool of modern engineering, with low-order finite element methods and assembled sparse matrices representing the industry standard for implicit analysis.
  We use performance models and numerical experiments to demonstrate that high-order methods greatly reduce the costs to reach engineering tolerances while enabling effective use of GPUs\revised{; these data structures also offer up to 2x benefit for linear elements}{}.
  We demonstrate the reliability, efficiency, and scalability of matrix-free $p$-multigrid methods with algebraic multigrid coarse solvers through large deformation hyperelastic simulations of multiscale structures.
  We investigate accuracy, cost, and execution time on multi-node CPU and GPU systems for moderate to large models \revised{(millions to billions of degrees of freedom)}{} using AMD MI250X (OLCF Crusher), NVIDIA A100 (NERSC Perlmutter), and V100 (LLNL Lassen and OLCF Summit), resulting in order of magnitude efficiency improvements over a broad range of model properties and scales.
  We discuss efficient matrix-free representation of Jacobians and demonstrate how automatic differentiation enables rapid development of nonlinear material models without impacting debuggability and workflows targeting GPUs.
  \revised{The methods are broadly applicable and amenable to common workflows, presented here via open source libraries that encapsulate all GPU-specific aspects and are accessible to both new and legacy code, allowing application code to be GPU-oblivious without compromising end-to-end performance on GPUs.}{}
\end{abstract}

\begin{IEEEkeywords}
portable, scalable, implicit solvers, matrix-free, solid mechanics, HPC
\end{IEEEkeywords}


\section{Introduction}

Solid mechanics simulations provide vital information for many engineering applications, using a large amount of computational resources from workstation to supercomputing scales.
The industry standard for implicit analysis uses assembled sparse matrices with low-order elements, typically $Q_1$ hexahedral and $P_2$ tetrahedral elements \cite{abaqus-web, ansys-mechanical-web}, with the linear systems solved using sparse direct solvers, algebraic multigrid, or multilevel domain decomposition.
This approach has two fundamental inefficiencies: poor approximation accuracy per Degree of Freedom (DoF) and high computational and memory cost per DoF due to choice of data structures and algorithms.
High-order finite elements implemented in a matrix-free fashion with appropriate preconditioning strategies can overcome these inefficiencies.

Solid mechanics models invariably have many stress singularities due to boundary conditions and possible reentrant corners; therefore, $h$-refinement with any finite element basis order $p$ will converge at the same low-order of accuracy.
Typically, $hp$-adaptive methods \cite{babuvska1994p} are used to resolve these singularities and enable geometric convergence.
Such methods are available in niche commercial products such as StressCheck \cite{stresscheck-web} as well as open source finite element libraries \cite{bangerth2009data,frauenfelder2002coo}, but are rarely used in production computational engineering.
This is attributed to accuracy requirements and constant factors: a low-order discretization can usually reach engineering tolerances with a coarse enough mesh that the modeling and implementation complexity of $hp$-adaptive methods are not justifiable, despite their clear asymptotic benefit.

Non-adaptive high order finite elements reduce complexity by being drop-in substitutes for low order elements if one can mesh the geometry more coarsely.
Quadratic \cite{schneider2022large} and higher order \cite{duester2003pfem} elements are often shown to be more accurate per DoF for large deformation analysis despite the presence of singularities preventing any asymptotic benefit.
However, such methods are rarely used due to high computation and memory costs for assembly and solution of the linear systems.
Sparse matrices for high-order elements have more nonzero entries per DoF: a $Q_1$ hexahedral element contributes 27 nodes per row while $Q_2$ elements have an average of 64 nodes per row, so every sparse matrix-vector product is more than twice as expensive per DoF.
Note that vertex separator size stays constant in $h$ versus $p$ refinement to the same number of DoFs, so sparse direct solvers have the same size supernodes and thus asymptotic complexity \cite{george1994computer}, although the memory use and leaf cost increases.
Meanwhile, algebraic multigrid (AMG) setup costs increase due to sparse matrix-matrix products, and the resulting solvers are observed to converge more slowly for high-order discretizations \cite{davydov2020matrix,weber2015pmultigrid,sundar2015comparison}, even when using specialized methods \cite{heys2005algebraic}.
A practical alternative to algebraic multigrid is $p$-multigrid \cite{ronquist1987spectral}, which is observed to be robust for finite element discretizations on unstructured meshes \cite{sundar2015comparison,weber2015pmultigrid} and pairs naturally with efficient matrix-free data structures \cite{brown2021libceed}.

Krylov solvers and preconditioners rely on matrix-vector operations that perform two floating point operations (FLOPs) per stored nonzero.
For sparse matrix representations, each nonzero requires 12 bytes (or 16 if using 8-byte integers) to store double precision real values and their indices, yielding an arithmetic intensity \cite{williams2009roofline} of 1 FLOP/6 bytes.
Modern CPU and GPU hardware \cite{karlrupp_CPU-GPU-MIC} provides upwards of 10 FLOPs per byte of memory streamed from DRAM or GPU device memory, and thus iterative sparse linear solvers saturate memory bandwidth at less than 2\% of the device's floating point peak.
Fortunately, matrix-free methods \cite{kolev2021ecpceed,abdelfattah2021gpu,may2014ptatin} enable greatly reduced memory bandwidth requirements, often in exchange for modestly more FLOPs.
In this new performance model, equipped with matrix-free $p$-multigrid methods and AMG coarse solvers, high-order methods become cheaper per DoF than low-order methods (assembled or not), enabling significantly faster and cheaper simulations at engineering tolerances.

In this paper, we demonstrate that high order methods improve accuracy per DoF for hyperelastic simulations of multiscale structures, even at coarse tolerances in the presence of singularities.
We also demonstrate that such models can be solved robustly on a range of modern architectures at a fraction of the cost of per DoF of linear elements, using abstractions amenable to encapsulation and productive development.
These benefits are multiplicative, reducing the cost of implicit finite element analysis by up to an order of magnitude in terms applicable to existing analysis pipelines.
This paper is organized as follows: \autoref{sec:model} introduces the hyperelastic formulation and finite element representation, \autoref{sec:solver} describes the solver design and implementation, \autoref{sec:accuracy} investigates accuracy in terms of mesh resolution and number of DoFs, \autoref{sec:performance} investigates efficiency per DoF and solver robustness, and \autoref{sec:discussion} discusses implications and opportunities for further work.


\section{Mathematical Model}\label{sec:model}

\subsection{Variational Form for Hyperelasticity}

In hyperelasticity, one seeks the displacement field $\bm u(\bm X)$ expressing the current (deformed) configuration $\bm x = \bm X + \bm u$ in terms of the initial configuration $\bm X$.
An isotropic Neo-Hookean material is defined by its strain energy density
\begin{align}\label{strain-energy}
\psi \left( \bm e \right) &= \frac{\lambda}{2} \left( \log J \right)^2 -\mu \log J + \frac{\mu}{2} \left( \trace \bm b -3 \right) \nonumber \\
&= \frac{\lambda}{2} \left( \log J \right)^2 -\mu \log J + \mu \, \trace \bm e,
\end{align}
where $\bm b = (\nabla_X \bm x)(\nabla_X \bm x)^{T}$ is the left Cauchy-Green tensor, $J = \det \left( \nabla_X \bm x \right)$, $\mu$ and $\lambda$ are the Lam\'{e} parameters, and $\bm e$ is the Green-Euler strain tensor,
$$
\bm e \equiv \frac{1}{2}\left( \bm b -\bm I \right)= \frac{1}{2} \left( \nabla_X \bm u + \left( \nabla_X \bm u \right)^T + \left( \nabla_X \bm u \right) \left( \nabla_X \bm u \right)^T \right).
$$

For a domain $\Omega_0 \subset \mathbb{R}^3$ with boundary $\partial \Omega_0$ and the finite element space $\mathcal{V} \subset H^1 \left( \Omega_0 \right)$, the variational problem finds a solution $\bm u \in \mathcal{V}$ such that \cite{davydov2020matrix, mehraban2021efficient}

\begin{equation}\label{variational-form}
\underbrace{\int_{\Omega_0}{\nabla_{x} \bm v \tcolon \bm \tau}\, dV}_{a \left( \bm v, \bm u \right) }
 =\int_{\Omega_0}{\bm v \cdot \rho_0\bm g}\, dV
 + \int_{\partial\Omega_0}{\bm v \cdot \bar{\bm t}} \, dS  \quad \forall \bm v \in \mathcal{V},
\end{equation}
where $\rho_0$ is the initial mass density, $\bm g$ is the body force per unit mass, $\bar{\bm t}$ is the prescribed traction, $\nabla_x$ denotes spatial derivative with respect to the current configuration, and $\bm \tau$ is the Kirchhoff stress given by \cite{holzapfel2002nonlinear}
\begin{align}\label{Kirchhoff-stress}
\bm \tau =\frac{\partial \psi}{\partial \bm e} \bm b & = \mu \left( \bm b -\bm I \right) + \lambda \log J \, \bm I \nonumber \\
& =  2 \mu \, \bm e + \lambda \log J \, \bm I.
\end{align}

In order to solve \eqref{variational-form} using a Newton iteration algorithm, we need the Jacobian form of $a \left( \bm u, \bm v \right)$ as \cite{mehraban2021efficient}
\begin{equation}\label{Jacobian}
\diff a \left( \bm v, \diff \bm u; \bm u \right) = \int_{\Omega_0} \nabla_x  \bm v \tcolon \left( \diff \bm \tau - \bm \tau \left( \nabla_x \diff \bm u \right)^{T} \right) \, dV,
\end{equation}
where
\begin{multline}\label{dtau}
\diff \bm \tau - \bm \tau \, \left( \nabla_x \diff\bm u \right)^T = \nabla_x \, \diff\bm u \, \bm \tau + \lambda \trace \diff\bm \epsilon \, \bm I \\
+ 2 \left( \mu - \lambda \, \log\, J \right) \, \diff\bm \epsilon,
\end{multline}
with
\begin{equation}\label{depsilon}
\diff\bm \epsilon  = \dfrac{1}{2} \left( \nabla_x \, \diff\bm u  \, + \,  \left( \nabla_x \, \diff\bm u \right)^T \right).
\end{equation}

\subsection{Matrix-free Finite Element Formulation}

The residual \eqref{variational-form} and Jacobian \eqref{Jacobian} forms require derivatives with respect to (solution dependent) current configuration $\bm x$.
For efficient matrix-free discretization \cite{brown2010efficient,KnepleyBrownRuppSmith13}, we pull these forms back to reference coordinates $\bm\xi$ by way of the chain rule $\nabla_{x}(\cdot) = \frac{\partial(\cdot)}{\partial \bm \xi} \left[ \big( \frac{\partial \bm\xi}{\partial \bm X} \big) \big( \frac{\partial \bm X}{\partial \bm x} \big) \right]$, where the part in brackets will move into the variational forms evaluated at quadrature points and $\nabla_{\xi} = \frac{\partial}{\partial \bm\xi}$ is applied by batched element algebra.
We explain this approach for a general Dirichlet problem: find $\bm u \in \mathcal{V}_0$ such that
\begin{multline}\label{residual}
\langle \bm v, \bm f \left( \bm u \right) \rangle = \int_{\Omega_0} \left[ \bm v \cdot  \bm f_0 \left( \bm u, \nabla_x \bm u \right) \right. \\
+ \left. \nabla_x \bm v \tcolon \bm f_1 \left( \bm u, \nabla_x \bm u \right) \right] dV = 0, \quad \forall \bm v \in \mathcal{V}_0
\end{multline}
(cf. \eqref{variational-form} without traction, where $\bm f_0 = -\rho_0 \bm g$ and $\bm f_1 = \bm \tau$).
The discrete form of \eqref{residual} is given by
\begin{multline} \label{residual-discrete}
\bm F \left( \bm u \right) = \sum_{e} \left( \mathcal{E}^e \right)^T \left[ \textcolor{white}{\Bigg|} \left( \bm B_I^e \right)^T \bm W^e \Lambda \left( \bm f_0 \left( \bm u^e, \nabla_x \bm u^e \right) \right) \right. \\
+ \left. \sum_{i = 1}^{\text{dim}} \left( \bm B_{x, i}^e \right)^T \bm W^e \Lambda \left( \bm f_1 \left( \bm u^e, \nabla_x \bm u^e \right) \right) \right],
\end{multline}
where $\mathcal{E}_e$ is the element $e$ restriction operator that separates DoFs based on the elements they belong to, and $\Lambda$ represents pointwise function evaluation.
The diagonal weighting $\bm W^{e} = \det \left( \nabla_{\xi}\bm X \right) \Lambda \left( \hat W \otimes \hat W \otimes \hat W \right)$ are quadrature weights mapped to the physical element.
Both $\bm f_0$ and  $\bm f_1$ come from the constitutive law and its tangent where
$\bm u^e = \bm B_I^e {\mathcal{E}^e\bm u}$ and
$$\nabla_x \bm u^e = \left[ \bm B_{x, i}^e \left( \mathcal E^e \bm u \right) \right]_{i = 1}^{\text{dim}} = \sum_{j = 1}^{\text{dim}} \left[ \bm B_{\xi, j} \left( \mathcal{E}^e \bm u \right) \frac{\partial \bm \xi_j}{\partial \bm x} \right],$$
where $\bm u$ is the assembled solution vector, $\text{dim}$ the dimensionality of the problem (for our use cases $\text{dim}=3$), and
\begin{align}\label{tensorDense}
\bm B_I      &= B_I \otimes B_I \otimes B_I, &
\bm B_{\xi, 1} &= B_{\xi} \otimes B_I \otimes B_I, \nonumber\\
\bm B_{\xi, 2} &= B_I \otimes B_{\xi} \otimes B_I, &
\bm B_{\xi, 3} &= B_I \otimes B_I \otimes B_{\xi},
\end{align}
are evaluations on reference elements written in terms of their one dimensional tabulations $B_I$ and $B_{\xi}$ of shape functions and their derivatives at quadrature points.
The representation \eqref{tensorDense} implies nine tensor contractions to compute a gradient, but this can be reduced to six by first applying $\bm B_I$ (3 contractions) and then applying $\hat{\bm B}_{\xi, 1} = \left( B_{\xi} B_I^{\dagger} \right) \otimes I \otimes I$, and similarly for the other directional derivatives, where $B_I^{\dagger}$ is the pseudo-inverse satisfying $B_{I}^{\dagger}B_{I} = I$ because $B_{I}$ has full column rank.
Asymptotically fast structure can also be exploited for simplicial elements \cite{kirby2011fast,chan2017bernstein}, but the constants are large enough that direct assembly of the reference element gradient $\bm B_{\xi}$ is preferred for the modest basis order considered here.

Pulling \eqref{residual-discrete} back to reference coordinates, we have
\begin{multline} \label{residual-discrete-ref}
  \bm F \left( \bm u \right) = \sum_{e} \left( \mathcal{E}^e \right)^T \begin{bmatrix} \bm B_I \\ \bm B_\xi \end{bmatrix}^{T}
  \bm W^e \Lambda \begin{bmatrix} \left( \hat{\bm f}_0 \left(\bm u^e, \widehat{\nabla_{\xi} \bm u}^e \right) \right) \\
    \left( \hat{\bm f}_1 \left( \bm u^e, \widehat{\nabla_{\xi} \bm u}^e \right) \right) \end{bmatrix},
\end{multline}
where $\hat{\bm f}_{0} \left( \bm u^{e}, \widehat{\nabla_{\xi}\bm u^{e}} \right) = \bm f_{0} \left( \bm u^{e}, \widehat{\nabla_{\xi}\bm u}^{e} \nabla_{x}\bm\xi \right)$ and
$$
\hat{\bm f}_{1} \left( \bm u^{e}, \widehat{\nabla_{\xi}\bm u^{e}} \right) = \left( \nabla_{x}\bm\xi \right)^T \bm f_{1} \left( \bm u^{e}, \widehat{\nabla_{\xi}\bm u}^{e} \nabla_{x}\bm\xi \right).
$$
While this interface brings the isoparametric mapping into quadrature functions, the work outside these quadrature routines shares the same data and can be batched over elements $\widehat{\nabla_{\xi}\bm u}^{e} = \bm B_\xi \mathcal E^{e}\bm u$, leading to improved vectorization and data reuse.
Moreover, this abstraction provides ready access to element length measures (useful in stabilized methods for transport-dominated processes \cite{hughesetal2010}) and allows optimized data representations, such as bypassing initial configuration because \eqref{dtau} can be evaluated strictly in current configuration, a technique equivalent to that of \cite{davydov2020matrix}.

The Jacobian action can be computed \cite{brown2010efficient} similar to the residual \eqref{residual-discrete-ref},
\begin{equation} \label{Jacobian-discrete-ref}
  \bm J \diff\bm u = \sum_{e} \left( \mathcal{E}^e \right)^T \begin{bmatrix} \bm B_I \\ \bm B_\xi \end{bmatrix}^{T}
  \bm W^e \Lambda \begin{bmatrix}
\hat{\bm f}_{0,0} & \hat{\bm f}_{0,1} \\
\hat{\bm f}_{1,0} & \hat{\bm f}_{1,1}
\end{bmatrix}
\begin{bmatrix} \bm B_I \\ \bm B_\xi \end{bmatrix}
\mathcal E^{e} \diff\bm u
\end{equation}
where
\begin{align*}
  \hat{\bm f}_{i,0} &= \frac{\partial \hat{\bm f}_{i} \left( \bm u^{e}, \widehat{\nabla_{\xi}\bm u^{e}} \right)}{\partial \bm u^{e}}, &
  \hat{\bm f}_{i,1} &= \frac{\partial \hat{\bm f}_{i} \left( \bm u^{e}, \widehat{\nabla_{\xi}\bm u^{e}} \right)}{\partial \widehat{\nabla_{\xi}\bm u^{e}}}.
\end{align*}
Since functional derivatives commute with pull backs, one could equivalently differentiate $\bm f_{i}$ in physical space to produce $\bm f_{i,j}$, then pull back to $\hat{\bm f}_{{i,j}}$.

\subsection{libCEED Abstraction}

\begin{figure}
\includegraphics[width=.99\linewidth]{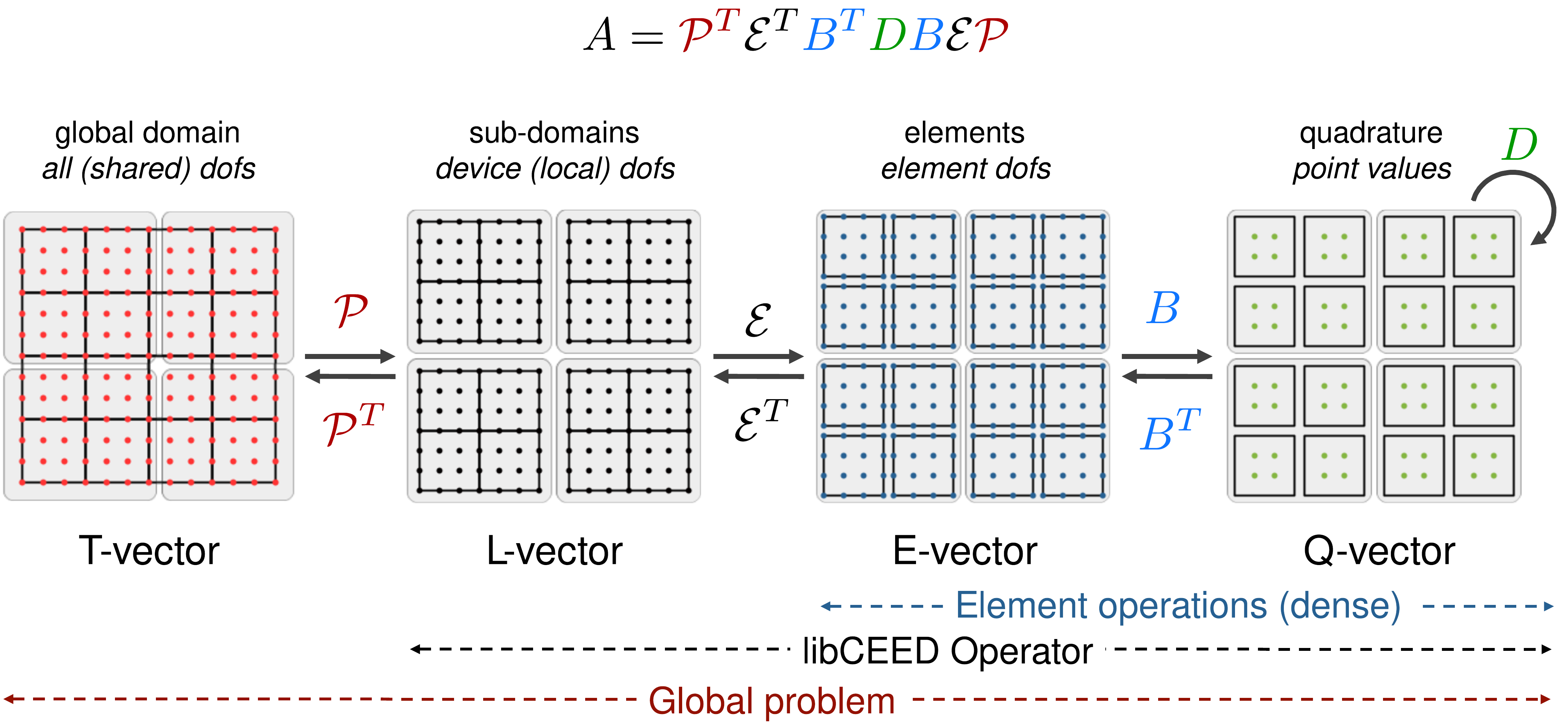}
\caption{libCEED composes local (L-vector to L-vector) operations from element restriction $\mathcal E$, basis $\bm B$, and quadrature-point functions $\bm D$.
  \revised{A T-vector represents the non-overlapping parallel partition of DoFs, as needed by nonlinear and linear algebraic solvers. The L-vector is localized per device (e.g., MPI rank or GPU context) with any ghost values replicated into each part. The E-vector (restricted to elements) and Q-vector (evaluated to quadrature points) exist only conceptually in our optimized implementation, since restriction $\mathcal E$, basis  $\mathcal B$, and user-provided quadrature function $\mathbf D$ are fused into one kernel.}{}
}\label{fig:libceedapi}
\end{figure}

Systems of equations with the form \eqref{residual-discrete-ref} and \eqref{Jacobian-discrete-ref} admit a natural implementation via libCEED \cite{brown2021libceed}, which provides fast algebra for element-based discretizations on CPUs and GPUs.
\autoref{fig:libceedapi} illustrates the action of an arbitrary finite element operator,
\begin{equation}
\mathbf{A} = \mathcal{P}^T \mathcal{E}^T \mathbf{B}^T \mathbf{D} \mathbf{B} \mathcal{E} \mathcal{P},
\label{libceed_representation}
\end{equation}
where $\mathcal{P}$ represents the parallel communication portion of the element restriction operator, $\mathcal{E}$ represents the local portion of the element restriction operator, $\mathbf{B}$ represents the basis action kernels that provide solution values and derivatives at the quadrature points given by $\mathbf{B}_I$ and $\mathbf{B}_\xi$, and $\mathbf{D}$ (which may be linear or nonlinear) represents the pointwise representation of the weak form, given by $\hat{\bm f}_i$ and $\hat{\bm f}_{i, j}$ as well as the element quadrature weights $\mathbf{W}$ and geometric factors $\nabla_{\xi}\bm X$.

\section{Solver Design}\label{sec:solver}

\subsection{Nonlinear and Iterative Solvers}\label{sec:iterative}
Large deformation solid mechanics exhibits both geometric and material nonlinearities, leading to path dependence by which there can be multiple static solutions for a specified set of boundary conditions.
To disambiguate the multiple solutions, we solve \eqref{residual-discrete} as a non-autonomous differential algebraic equation of index 1, with boundary conditions/loading a function of time $t \in [0, 1]$.
The examples in the present study use applied load (rather than displacement) and use backward Euler from the Portable, Extensible Toolkit for Scientific Computation (PETSc) \cite{petsc-user-ref, mills2021ecppetsc}, with extrapolation-based hot starts disabled for simplicity.
Each pseudo time step requires a nonlinear solve, which is implemented using PETSc's Scalable Nonlinear Equations Solver (SNES).
We consider Newton-CG and L-BFGS methods in which a multigrid V-cycle is used either as a preconditioner for conjugate gradients or as the ``initial inverse Hessian'' scaling for L-BFGS \cite{brown2013quasinewton}.
In both cases, we use a ``critical point'' line search, which supposes that the residual is the functional gradient of a latent objective function, $\bm F(\bm u) = \nabla_{\bm u}\Psi(\bm u)$ and uses one step of a secant method to find $\alpha$ for which $\bm F(\bm u + \alpha \diff \bm u)^{T} \diff \bm u = 0$\revised{,}{} where $\diff \bm u$ is the search direction found by Newton or L-BFGS.
This line search is inspired by the strong Wolfe conditions in optimization \cite{NW99}, but without explicit evaluation of the objective $\Psi$, which may not be available or may not exist (e.g., for non-conservative models).

The linear solve and multigrid preconditioner uses PETSc's Krylov Subspace (KSP) and Preconditioning (PC) tools.
When using Newton-CG, each Newton step $\bm J \diff \bm u = - \bm F(\bm u)$ is solved to a relative tolerance of $10^{-3}$ in the natural norm.
To clarify preconditioner robustness, we report condition number estimates for the preconditioned operator obtained from the tridiagonal matrix implied by the CG/Lanczos recurrence, with similar estimates of the maximum eigenvalue used in the Chebyshev smoothers.

\subsection{Matrix-free \texorpdfstring{$p$}{\textit{p}}-multigrid}

Multigrid methods provide an efficient preconditioning framework for obtaining uniform convergence rates with respect to resolution and model extent.
$p$-type multigrid, developed by Ronquist and Patera \cite{ronquist1987spectral}, is \textit{sensibly independent} of the number of elements and polynomial order of the element bases.
In $p$-multigrid, the discretization is coarsened by reducing the polynomial order of the basis functions, in contrast to $h$-multigrid, where the mesh is coarsened by aggregating elements.
$p$-multigrid is a natural fit for high-order finite elements on unstructured meshes and can be implemented with operators represented in libCEED's computationally efficient form in \eqref{libceed_representation}.

\begin{figure}
~\\
\begin{algorithmic}[1]
\State Compute $\bm{u}_k$
\State $\bm{u}_k \gets \bm{u}_k + \hat{\bm{M}}^{-1} \left( \bm{b} - \bm{A}_f \bm{u}_k \right)$ \Comment{pre-smooth}
\State $\bm{r} = \bm{P}_{\text{ctof}}^T \left( \bm{b} - \bm{A}_f \bm{u}_k \right)$             \Comment{restrict the residual}
\State $\bm{A}_c \bm{e} = \bm{r}$                                                              \Comment{Solve on coarse grid}
\State $\bm{u}_k \gets \bm{u}_k + \bm{P}_{\text{ctof}} \bm{e}$                                 \Comment{prolong error correction}
\State $\bm{u}_k \gets \bm{u}_k + \hat{\bm{M}}^{-1} \left( \bm{b} - \bm{A}_f \bm{u}_k \right)$ \Comment{post-smooth}
\end{algorithmic}
\caption{The multigrid algorithm is applied recursively, with Galerkin coarse operator $\bm{A}_{c} = \bm{P}_{\text{ctof}}^{T}\bm{A}_{f}\bm{P}_{\text{ctof}}$ constructed matrix-free until coarsening to linear elements, then via algebraic multigrid.}\label{multigrid_algorithm}
\end{figure}
\autoref{multigrid_algorithm} describes a standard V-cycle \cite{brandt1982guide}.
In this algorithm, $\bm{A}_f$ is the operator on the fine grid, $\bm{P}_{\text{ctof}}$ is the coarse to fine grid prolongation operator, and $\hat{\mathbf{M}}$ a separate preconditioner used for smoothing.
We use the transpose of the prolongation operator as the fine to coarse grid restriction operator to preserve symmetry and prevent aliasing.
We define \revised{the}{} smoother (and implicitly, $\hat{\bm M}$) as the 2nd order Chebyshev/Jacobi iteration targeting the range $[0.1 \lambda_{\max}, 1.1 \lambda_{\max}]$, where $\lambda_{\max}$ is the eigenvalue estimate computed by 10 Lanczos iterations applied to a ``rough'' seed vector during preconditioner setup.
Prolongation is expressed within the libCEED abstraction of \autoref{fig:libceedapi} via
\begin{equation}
\bm{P}_f^p = \sum_e \left( \mathcal{E}_f^e \right) ^T \Lambda \left( m_f^{-1} \right) \bm{B}_{\text{ctof}}^e \mathcal{E}^e_c,
\label{mg_prolong}
\end{equation}
where $\bm{B}_{\text{ctof}}^e$ is the interpolation kernel from the lower order to the higher order finite element, defined by \eqref{tensorDense}, and $m_f = \mathcal{E}_f \mathcal{E}_f^T \mathcal{E}_f$\revised{}{1} is a pointwise scaling factor for the multiplicity of nodes shared between elements on the fine grid.

The Jacobian on each level is represented by the 17 scalar values per quadrature point, $\nabla_{x}\bm\xi, \bm \tau, \log J$, and the quadrature weight, as used in \eqref{dtau} and \eqref{Jacobian-discrete-ref}; cf. \autoref{table:different-storage}.
Coarse level discretizations are defined using the same quadrature points and Jacobian representation with coarser basis functions, which is an exact Galerkin method.
An alternative needing somewhat more memory, but less memory bandwidth to apply coarse operators, would be to rediscretize by re-evaluating the nonlinearity on a smaller set of quadrature points (sufficient for the lower order polynomials of the coarse space).
Note that coarsening from $Q_{2}$ to $Q_{1}$ elements reduces the number of DoFs by a factor of 8 and reduces the number of nonzeros per row \revised{}{from} (asymptotically on 3D models) by a factor of $64/27$.
When using direct solvers in 3D, this reduces the vertex separator by a factor of 4 and thus supernode factorization (the asymptotically dominant cost) by a factor of 64.
When using algebraic multigrid, this reduces the number of nonzeros in an assembled matrix by nearly 20 times (thereby reducing AMG setup and smoothing cost) as well as typically improving convergence rates.

\subsection{Portability and productivity}

\begin{figure}
  \centering
  \includegraphics[width=.9\linewidth]{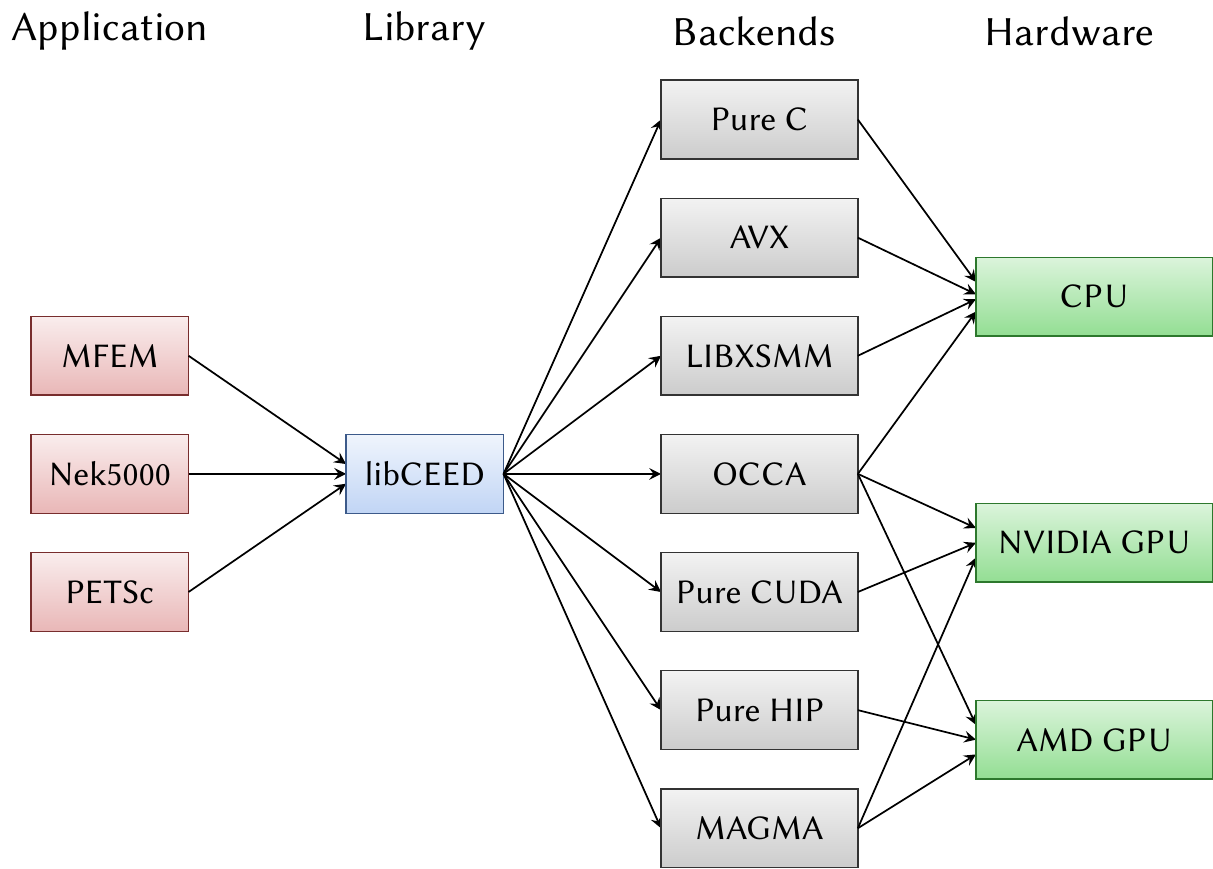}
  \caption{libCEED Backends}\label{fig:libceedbackends}
\end{figure}

PDE-based models contain symmetry/conservation structure, which is subject to change by a limited subset of stakeholders, and material models (extending \eqref{Kirchhoff-stress}) requiring frequent extensions by scientists and engineers who are not sophisticated numerical analysts or software developers.
It is thus important that materials developers have a simple, debuggable environment for development and testing.
libCEED \cite{brown2021libceed} provides fast algebra for element-based computation on CPUs and GPUs, meant for easy embedding in existing applications and enabling high performance on multiple architectures (\autoref{fig:libceedbackends}\revised{)}{} from a single source code, with run-time selection of the backend.

The CPU backends call conventionally compiled functions to apply residuals \eqref{residual-discrete-ref} and Jacobians \eqref{Jacobian-discrete-ref} at quadrature points, thereby enabling a rich debugging experience.
CPU backends implement the element action $\bm B$ using tensor contractions with architecture-specific vectorization (e.g., AVX intrinsics, LIBXSMM \cite{heinecke2016libxsmm}).
The GPU backends used in these experiments create a fused kernel containing the entire $\mathcal{E}^T \bm{B}^T \bm{D} \bm{B} \mathcal{E} \bm u$ of \eqref{libceed_representation}.
The source code for the application of the weak form at the quadrature points, $\bm{D}$, is transformed into an appropriate CUDA or HIP device function to be called as part of the fused operator.
The resulting kernel is compiled at runtime via NVRTC/hipRTC, inlining the above call and making loop bounds and memory access offsets compilation constants, thereby improving register allocation and performance.
\revised{In this implementation, there is no difference in user code to run on the different architectures.
  Indeed, the target CPU or (AMD or NVIDIA) GPU may be selected at run-time and need not be uniform across an application.
  All architecture-specific code is contained within libCEED backends and within PETSc and hypre \cite{falgout2021porting} numerical kernels that are entirely independent from the application.}{}

\subsection{Parallelism and GPUs}

While libCEED provides fast algebra on individual CPUs and GPUs, it is important that all problem-sized data stay resident on GPUs thoughout the parallel solve.
PETSc provides matrix and vector operations on the GPU, including the Galerkin product $\bm{P}_{\text{ctof}}^{T}\bm{A}_{f}\bm{P}_{\text{ctof}}$ in algebraic multigrid setup, with ability to use external libraries like Kokkos \cite{trott2022kokkos} and hypre \cite{falgout2021porting} as well as CUDA and ROCm vendor libraries.
Communication and computation are overlapped where possible, with message packing taking place on the GPU along with persistent nonblocking sends and receives using GPU-aware MPI, all via the ``star forest'' \cite{zhang2022petscsf} abstraction.

Compressed sparse row (CSR) type matrices are desirable for algebraic multigrid setup and solves, and have historically been created by preallocating and adding values to logically dense blocks using PETSc's \code{MatSetValues}, typically one block per element in a finite element computation.
This interface keeps memory utilization low, but is too fine-grained for efficient computing on GPUs and requires a binary search to find the insertion location in the CSR matrix.

We have developed a new interface in PETSc, based on a split-phase COO specification that enables efficient GPU assembly with strong encapsulation.
Previous literature \cite{fu2014architecting,cecka2011assembly,markall2013finite,dziekonski2012finite} on GPU-based sparse matrix assembly, including those using COO format \cite{dziekonski2012finite}, used coloring, atomics, or avoid assembling the global matrix to get around data races related to multiple finite elements summing into the same nonzero entries.
PETSc\revised{'s}{} new COO-based assembly avoids data races or atomics completely with new algorithms, and handles MPI parallelism.
The classic COO format consists of three arrays, \texttt{row[], col[], val[]}, of equal length, in which the assembled matrix $A$ is defined as the sum of each contribution \texttt{val[k]} to entry $a_{\texttt{row[k]},\texttt{col[k]}}$.
It is common in nonlinear and transient PDE solves that one needs to assemble a matrix with the same nonzero pattern but different numeric values.
PETSc's interfaces splits COO assembly into a symbolic \code{MatSetPreallocationCOO} in which the \texttt{row[], col[]} parts are provided, followed by one or more calls to \code{MatSetValuesCOO} in which the numeric array \texttt{val[]} is provided on-device.

In \code{MatSetPreallocationCOO}, which is done on\revised{-}{}host, we analyze the coordinates, exchange information about remote entries, finalize the sparsity pattern of diagonal and off-diagonal blocks, and preallocate memory for them on\revised{-}{}device.
This phase prepares to ignore negative indices (convenient for boundary conditions) and sum duplicate entries, as well as planning how to send remote entries to their destination, including which entries in \texttt{val[]} should be packed into send buffers.
In both PETSc's native and GPU formats as well as hypre's ParCSR, matrices are distributed row-wise across processes with diagonal (intra-process coupling) and off-diagonal (inter-process coupling) blocks stored separately in CSR format.
The arrays \texttt{row[], col[]} can be freed after the planning stage.

The \texttt{val[]} array is populated on-device using a libCEED kernel that performs the $\bm B^T \bm D \bm B$ portion of the operator \eqref{libceed_representation} (cf. \cite{KnepleyRuppTerrel2016}) as a triple matrix product, formulated such that each thread accumulates contributions for a particular element-based non-zero without forming an intermediate matrix.
When the number of basis nodes per element is low (up to and including $Q_2$ hexahedra), a two-dimensional thread block processes the row and column combinations in an element's output matrix; when this design would exceed the allowed number of threads per block, the assembly switches to a one-dimensional thread block with an additional loop in the kernel.
Runtime compilation through NVRTC/hipRTC ensures that all loop bounds are compile-time constants.
Each accumulated value is then assigned to the \texttt{val[]} array at a specified index determined by element and component ordering, and the final array is provided to \code{MatSetValuesCOO}.
Each entry (with nonnegative indices) is destined for the owned diagonal, owned off-diagonal block, or send buffer.
The implementation first calls a kernel to fill the send buffer and initiate the MPI communication, two asynchronous kernels for nonzeros in the diagonal and off-diagonal blocks, in which each thread accumulates into a single nonzero, and after completing communication, two similar kernels unpack entries from the receive buffer.



\section{Accuracy}\label{sec:accuracy}
\subsection{Pareto optimality at engineering tolerances}

\begin{figure}
\centerline{\includegraphics[width=.45\textwidth]{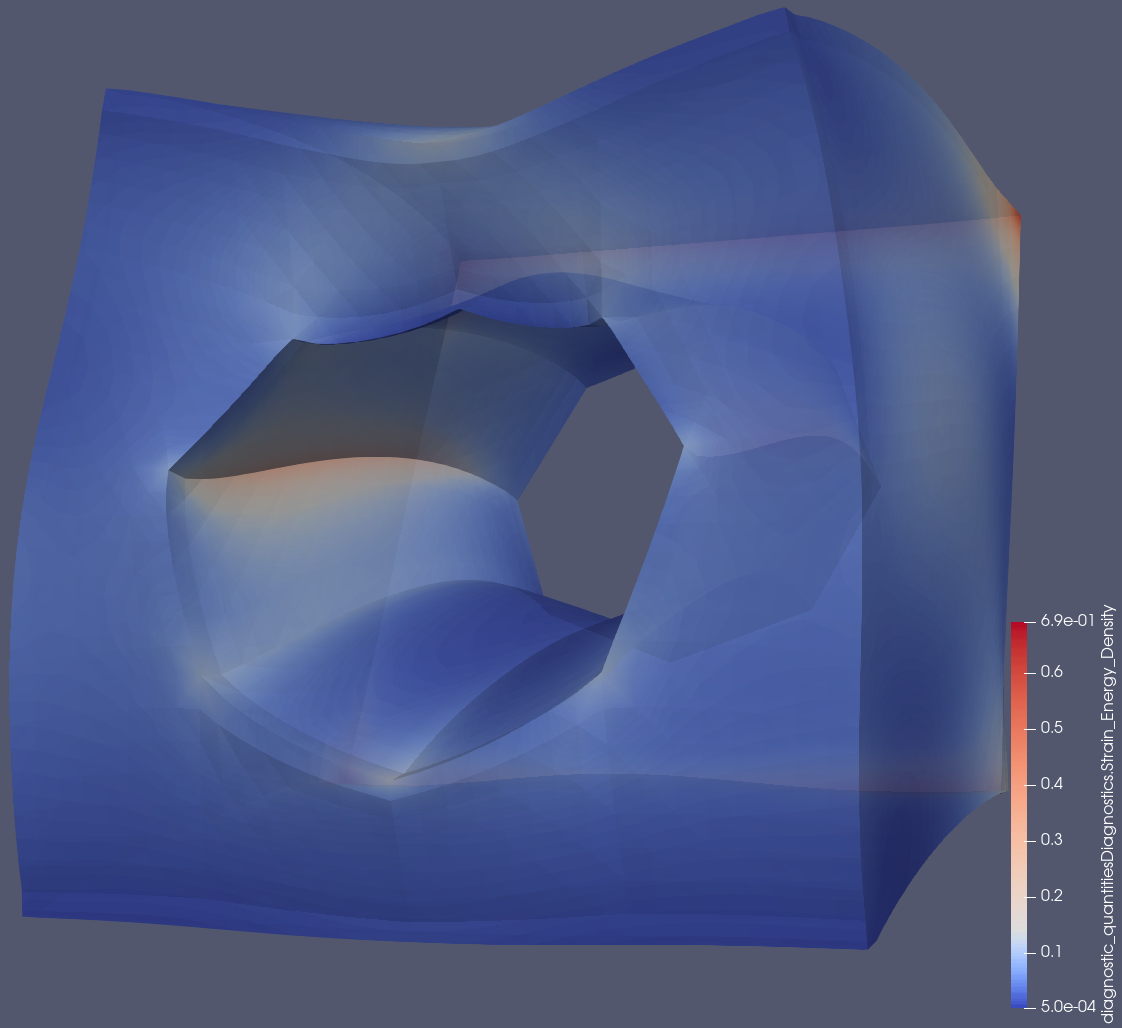}}
\caption{Visualization of the deformed state and strain energy singularities on \code{mesh A} of \autoref{fig:accuracy_refinement_study}, refined 3 times by splitting each hexahedron in 8 without snapping to geometry, and solved using $Q_{2}$ finite elements. There are physical singularities on the back surface (e.g., top-right corner) and non-physical singularities at the weak reentrant corners of the hole (which do not exist in the smooth model with exact cylinder).}
\label{fig:hole-strain-energy}
\end{figure}

Real-world structural mechanics problems have numerous reentrant corners and Dirichlet (fixed/clamped) to Neumann (free or applied traction) boundary condition transitions, each of which result in stress singularities.
Geometric convergence can be attained for such problems using $hp$-adaptive finite element methods \cite{babuvska1994p}, but such methods are rare in industrial practice because adequate tolerances can be achieved on coarser meshes.
This can be because the quantity of interest is not so sensitive or because unresolved features (beveling or bolts/washers) or physical yielding will alleviate the singularity in quantities of interest such as the von Mises stress.
Using high order finite elements on coarse meshes with singularities exposes some nuance, which we explore by way of a representative example.
Consider a unit cube with radius 0.3 cylindrical hole, fixed to a rigid boundary on one end and with applied tangential traction on the other.
\autoref{fig:hole-strain-energy} shows the deformed state and strain energy function for a Neo-Hookean material with Young's modulus 2.4 and Poisson ratio\revised{}{n} 0.4, and applied traction of 0.2.

\begin{figure}
	\centerline{\includegraphics[width=\linewidth]{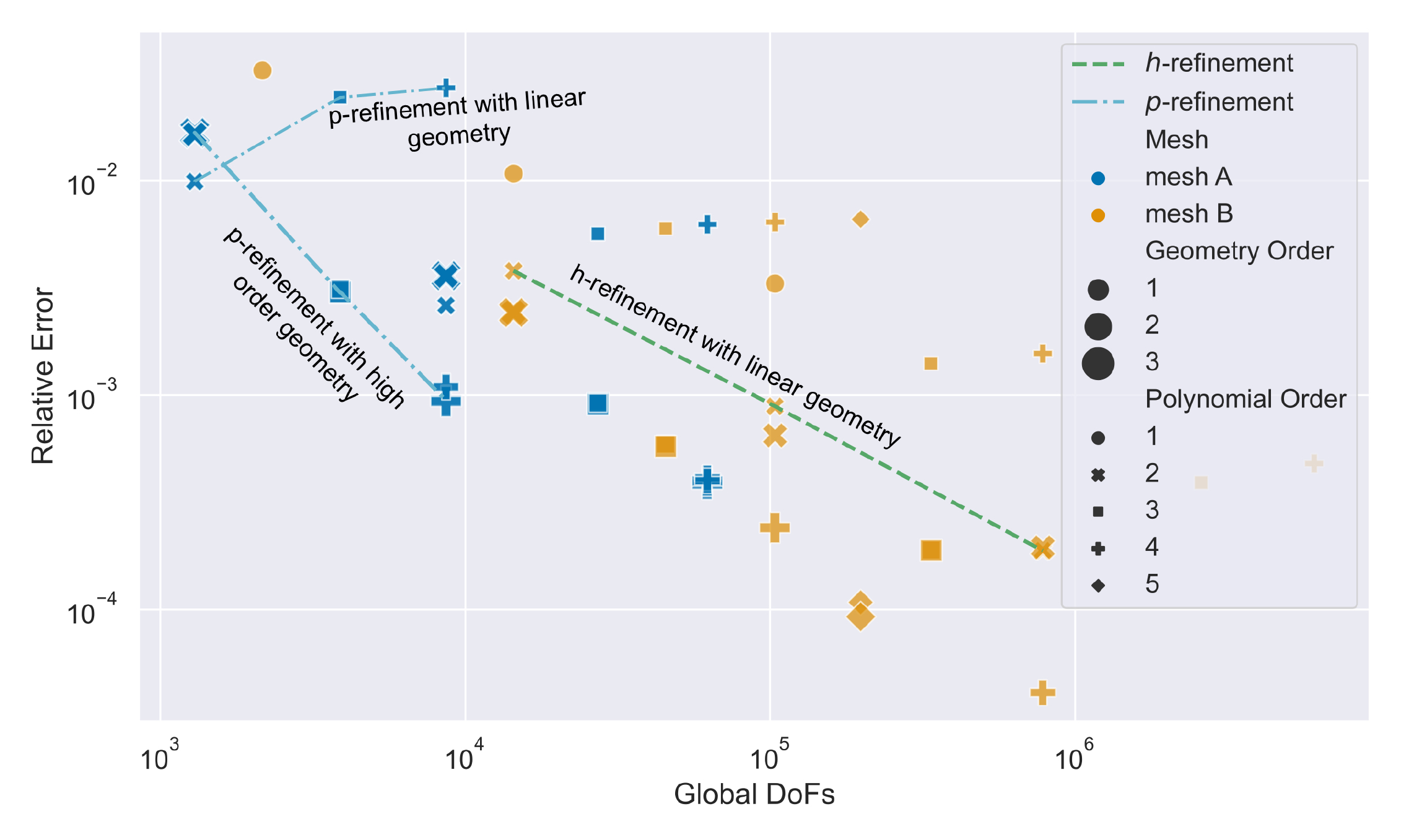}}
	\caption{Accuracy study showing relative error in total strain energy $\Psi$ versus DoFs for the bending experiment \autoref{fig:hole-strain-energy} under both $h$ refinement (same shape) and $p$ refinement (same color) with low and high order geometry. The Pareto front is toward the lower left and we observe that $h$ refinement always moves away from optimality.
  The slope of $h$ refinement is the same for all meshes and solution orders.
  $p$ refinement is very efficient so long as the geometry is at least quadratic, but causes errors to increase when $p$ refining on linear geometry due to resolution of the non-physical singularities.}
	\label{fig:accuracy_refinement_study}
\end{figure}

We perform a convergence study using linear and high-order geometry meshes produced by Gmsh \cite{geuzaine2009gmsh}, which can generate arbitrary order curved meshes.
\autoref{fig:accuracy_refinement_study} shows \revised{the}{} relative error in predicted total strain energy $\Psi$ (reference value computed on a highly-resolved mesh) versus DoFs for $h$ and $p$ refinement of the 36-element (3 layers deep) mesh evident in \autoref{fig:hole-strain-energy} (\code{mesh A}) as well as a more resolved \code{mesh B}.
We observe that $p$ refinement of very coarse meshes is the most efficient path to accuracy.
Our experience is consistent with prior empirical studies \cite{luo2002pversion} that a quadratic solution space can be paired with linear geometry, but we also find that further $p$ refinement is actively harmful as non-physical singularities are resolved.
\revised{}{When high order elements are used on the coarsest possible meshes (one element thick), the number of DoFs is often an order of magnitude less than would be required of linear elements to achieve the same accuracy.}

\revised{When high order elements are used on the coarsest possible meshes (one element thick), the number of DoFs is often an order of magnitude less than would be required of linear elements to achieve the same accuracy.}{}
Therefore, high order methods have better accuracy constants (thus favoring $p$ refinement), and yet they are rarely used in practice, mainly because the assembly and linear algebra are so much more expensive (no improvement in asymptotics).
Specifically, the FLOPs per DoF of naive matrix assembly for order $p$ polynomial basis functions in $d$ dimensions scales as $(p+1)^{2d}$ (this can be reduced by specialized methods \cite{melenk2001fully}) and the number of nonzeros per DoF in the assembled matrix scales with $p^{d}$.
The former was a historical bottleneck while the latter is fundamental given the high relative expense of data motion on modern hardware.
In contrast, when applying operators matrix-free with quadrature-point data, the storage per DoF declines (and is asymptotically constant) with increasing order $p$.
In the following sections, we show that solve costs \emph{decrease} with increasing $p$ using matrix-free $p$-multigrid and thus \autoref{fig:accuracy_refinement_study} is in fact \emph{generous} to the low-order methods.

\subsection{Schwarz Primitive extrusions under load}
Volumetric extrusions of triply periodic minimal surfaces have garnered interest during the additive manufacturing revolution for a range of applications from tissue membranes \cite{kapfer2011tissue} to metallurgy \cite{aboulkhair2019aluminium}.
We consider the Schwarz Primitive surface, which exhibits interesting geometric and material nonlinearities.
Prior finite element analysis of such models \cite{maskery2018insights} using voxelized meshes \cite{maskery2022flattpack} found that about 30k low order (Abaqus C3D8R) elements were needed to achieve an engineering tolerance of 1\%.
We consider conformal meshes that attain comparable accuracy with fewer DoF and much fewer elements.
To generate such meshes, start with a 24-element 2D manifold mesh of a single unit cell embedded in 3D, replicated to the prescribed extent in each embedding dimension.
This mesh is partitioned and distributed using ParMETIS, then refined with new nodes projected to the closest point on the implicit surface
$$ \cos 2\pi x + \cos 2 \pi y + \cos 2 \pi z = 0 .$$

\begin{figure}
  \centering
  \includegraphics[width=.45\textwidth]{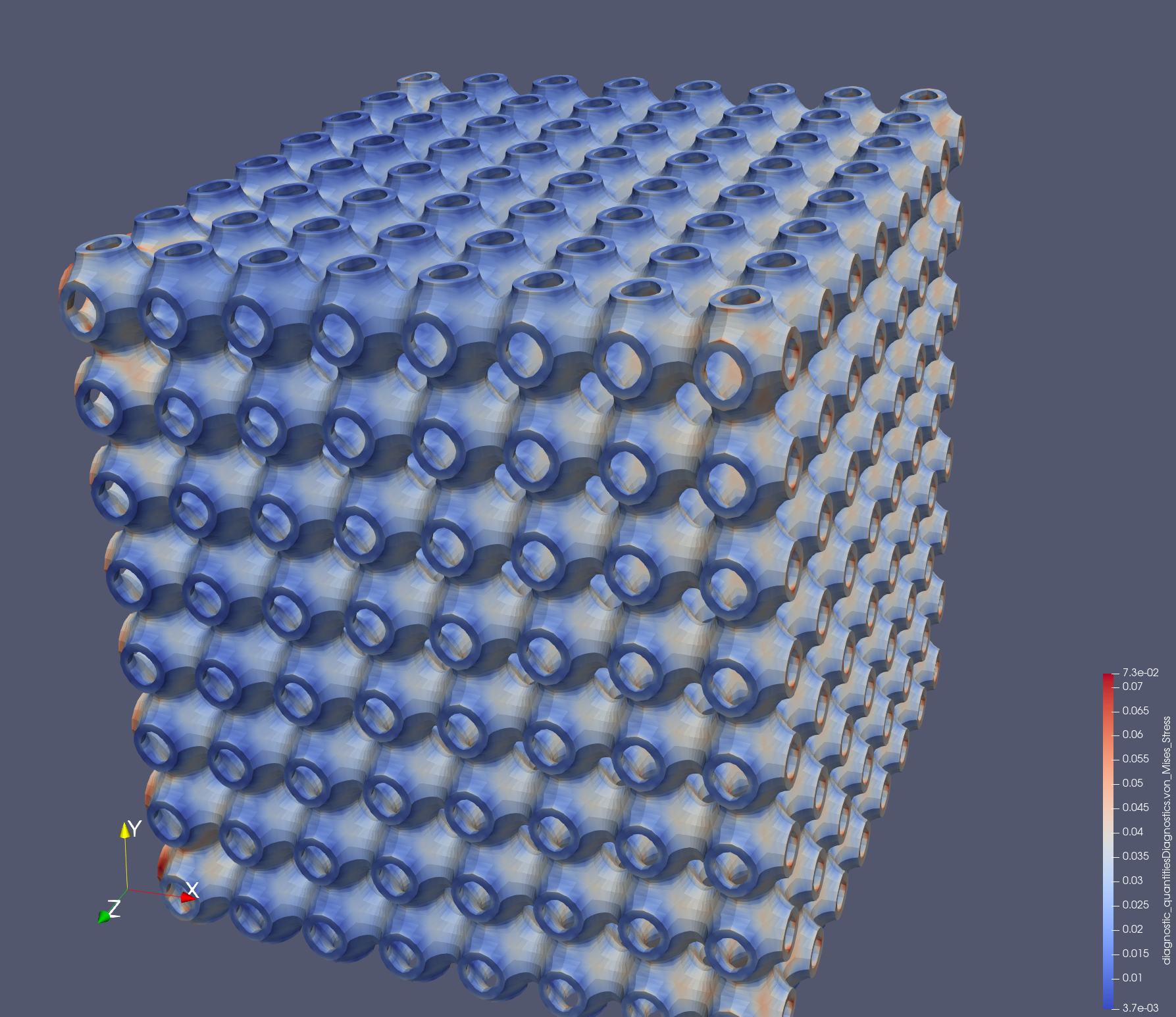}
  \caption{Extruded Schwarz Primitive surface under 12\% compressive strain, colored by von Mises stress. The left wall is fixed and a compressive force is applied to the facing surfaces on the right. The simulation used 2 refinements, 2 layers, thickness 0.2, and $Q_{2}$ elements.}
  \label{fig:schwarz-p}
\end{figure}

The resulting manifold mesh is extruded normal to this surface to the prescribed thickness and number of layers.
\autoref{fig:schwarz-p} shows such a model loaded to about 12\% strain on an extent $(8,8,8)$ model with about 11.8 million DoF (MDoF).
Larger and smaller models are created by changing the extent, keeping the applied surface traction constant so the deformation is similar.
These models, which are available in PETSc-3.17, provide excellent tests for solvers since they exhibit all compressive and bending modes, nonlinearities are activated at local and global scale, coarsening is inherently unstructured, and scaling is done by making the domain larger while achieving the same accuracy tolerances, in contrast to the common practice of refining a simpler domain to achieve unrealistically tight accuracy tolerances.

\begin{table}
  \renewcommand{\arraystretch}{1.3}
  \caption{Relative errors in maximum X and Y displacement and strain energy for a thickness 0.2 Primitive extrusion with extent $(4,3,3)$. Percent errors are calculated with respect to a reference (Order 2, Refinement 4, Layers 5). Errors under 5\% \revised{for}{} all 3 metrics are italicized. Configurations used in later studies are bolded. MDoFs are provided for comparison.}
  \label{table:schwarz_extrusion}
  \centering
  \begin{tabular}{rrrrrr D{.}{.}{2}}
    \toprule
           &         &         & \multicolumn{3}{c}{\% Error}
                                  \\\cmidrule(lr){4-6}
    Order &  Refinement &  Layers &  Disp. X &  Disp. Y &  Strain & \multicolumn{1}{r}{MDoF} \\
    \midrule
    3 &           3 &       1 & \textit{0.42} & \textit{0.47} & \textit{0.90} & 6.0 \\ 
    3 &           2 &       1 & \textbf{1.60} & \textbf{1.09} & \textbf{3.96} & 1.5 \\ 
    3 &           1 &       1 &            1.31 &            2.63 &          10.90 & 0.38 \\
    2 &           3 &       5 & \textit{0.42} & \textit{0.33} & \textit{0.92} & 7.3 \\ 
    2 &           3 &       2 & \textit{0.60} & \textit{0.77} & \textit{1.07} & 3.3\\ 
    2 &           2 &       2 & \textbf{2.54} & \textbf{2.64} & \textbf{4.82} & 0.84\\ 
    2 &           2 &       1 &            3.85 &            4.34 &           6.15 & 0.50 \\
    1 &           4 &       5 & \textit{2.15} & \textit{2.94} & \textit{2.23} & 4.0 \\ 
    1 &           3 &       5 &            6.67 &            8.66 &           7.36 & 1.0\\
    1 &           3 &       2 &            8.88 &           11.52 &           9.70 & 0.50 \\
    1 &           2 &       2 &           22.62 &           27.80 &          25.22 & 0.12 \\
    \bottomrule
    \end{tabular}
\end{table}

\autoref{table:schwarz_extrusion} quantifies the effect of mesh refinement and number of layers on the accuracy of a solve with linear, $Q_2$, and $Q_3$ elements, all with linear geometry.
We see that the $Q_{2}$ and $Q_{3}$ solutions are more accurate per DoF than those with linear elements.
Increasing the number of layers in the mesh helps decrease the relative error of the simulation; however, 1 or 2 layers is sufficient for both $Q_2$ and $Q_3$ elements to give solutions within typical engineering accuracy tolerances.

\section{Performance}\label{sec:performance}
\subsection{Compute environments}

We present GPU-based results on LLNL's Lassen, OLCF's Summit and Crusher, and NERSC's Perlmutter.
Lassen and Summit are both IBM POWER9 machines with 4 and 6 NVIDIA V100-SXM2 16 GiB GPUs per node, respectively.
Crusher is an early-access machine with the same node architecture as the upcoming Frontier.
Each node has one 64-core AMD EPYC 7A53 CPU and connected via Infinity fabric to 4 AMD MI250X GPUs, each of which consists of two GCDs that appear as logically separate GPUs with 64 GiB each.
The GCDs and GPUs are connected via high-bandwidth Infinity fabric, and 4 Cray network interfaces are connected directly to the 4 (dual GCD) GPUs.
Perlmutter, which is presently in early access, consists of nodes with one AMD EPYC 7763 CPU connected via PCIe-4.0 to 4 NVIDIA A100 40 GiB GPUs.
The GPUs are connected to each other with NVLink-3 and each node has 2 Cray network interfaces connected to the CPU.
To compare performance on these machines, we present achieved throughput (DoF/second) normalized by logical GPUs (each of which has similar power requirements).
GPU-aware MPI was used on Lassen (Spectrum MPI) and Crusher (Cray MPI), but was disabled on Summit (Spectrum MPI; because results were slower) and Perlmutter (Cray MPI; because of bugs). 
\autoref{table:compilers} describes the environment used on each machine.

\begin{table}[!t]
  \renewcommand{\arraystretch}{1.3}
  \caption{Environment for each HPC machine.}
  \label{table:compilers}
  \centering
  \begin{tabular}{c p{.75\linewidth}}
    \toprule
    Computer & Modules and environment \\
    \midrule
    Crusher & cce/13.0.0, rocm/4.5.2, craype-accel-amd-gfx90a \\
    Summit & gcc/9.3.0, cuda/11.1.1 \\
    Lassen & clang/13.0.1-gcc-8.3.1, cuda/11.2.0 \\
    Perlmutter & gcc/11.2.0, cuda/11.5.0, cray-mpich/8.1.13, craype-accel-nvidia90, cpe-cuda \\
    \bottomrule
  \end{tabular}
\end{table}

This study used the open source packages PETSc-3.17 \cite{petsc-user-ref}, hypre-2.24 \cite{falgout2021porting,baker2010elasticity}, Kokkos-3.6 \cite{trott2022kokkos}, ParMETIS 4.0.3 \cite{karypiskumar98}, libCEED-0.10.1 \cite{brown2021libceed}, and Ratel-0.1 \cite{ratel-web}.
The numerical experiments ``preload'' by doing a crude tolerance solve that is discarded before starting timers in order to provide consistent timing representative of longer-running simulations.
For more accurate profiling of individual events, the profiled runs include some unnecessary synchronization with the GPU, introducing a slight latency penalty to the smallest model sizes.

\subsection{Operator application efficiency}

\begin{figure}
  \centering
  \includegraphics[width=.45\textwidth]{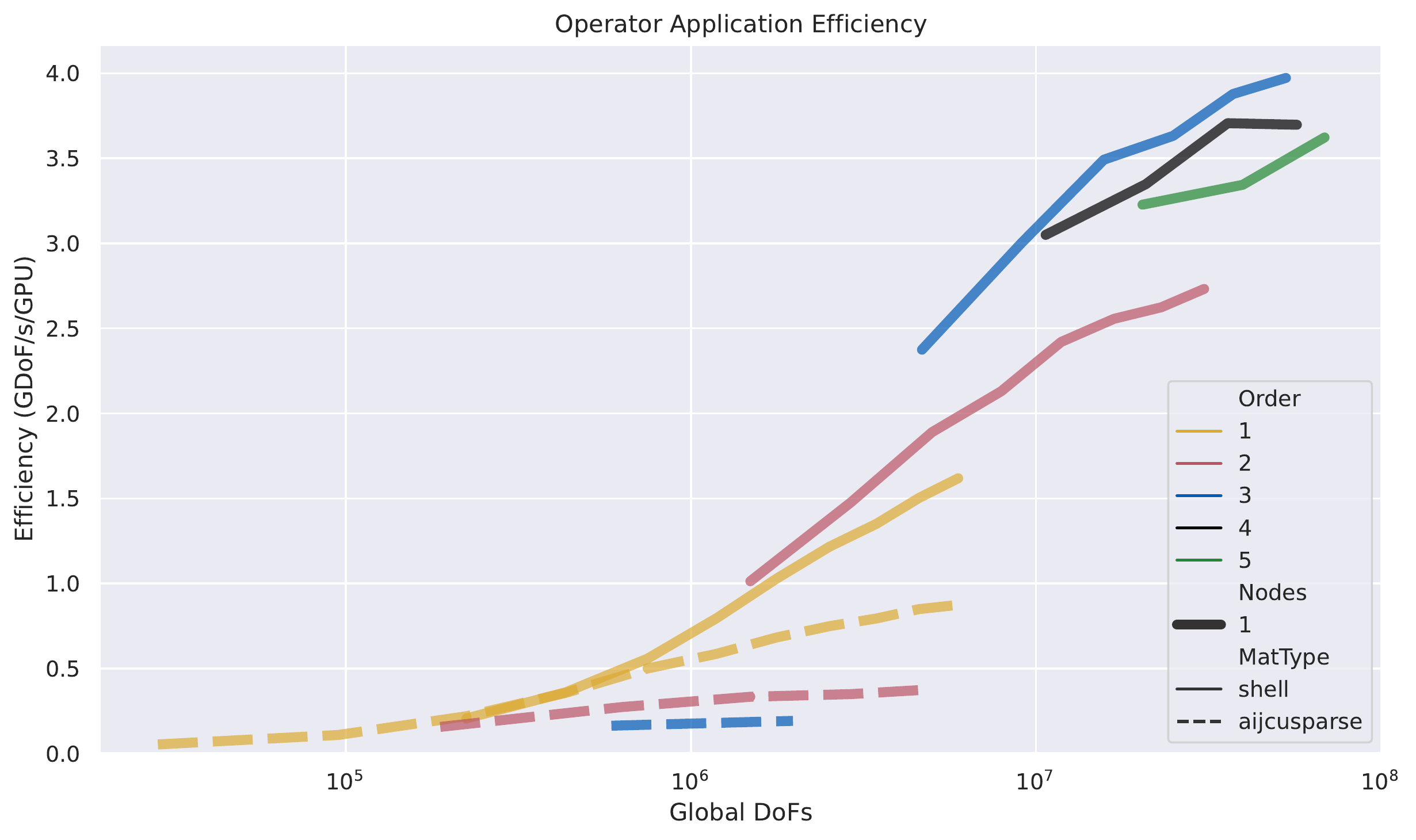}
  \caption{Parallel operator application efficiency running on Lassen with assembled \code{aijcusparse} and matrix-free \code{shell} operator representations.
	Note that \code{shell} becomes more efficient as the order increases while \code{aijcusparse} becomes less efficient.
	Both are latency-limited for smaller problem sizes (left side of the figure) and plateau as memory is filled for larger sizes.
	The \code{aijcusparse} cases run out of memory for smaller numbers of DoFs because high order methods yield many nonzeros per row.}
  \label{fig:op-apply}
\end{figure}

In addition to using less memory, the matrix-free representation is much more efficient per DoF to apply.
\autoref{fig:op-apply} presents performance by varying the domain size (thus total number of DoFs) on a test that runs 3 Newton steps with 500 iterations of CG per step (preconditioned by Jacobi).
The figure reports timing for the matrix multiplication operation only.
The assembled matrix for this model averages about 63 nonzeros per row (i.e., per DoF) and the empirical STREAM bandwidth on Lassen is $\qty{820}{GB\per\second}$ so we expect the matrix multiply to plateau at $820 / (63 \cdot 12) \approx \GDoFs{1}$ if it was only streaming \code{float64} matrix entries and \code{int32} column indices, without any cost to communicate or access vectors.
We see that it achieves nearly that and similar math shows that the high order discretizations (with more nonzeros per DoF) also nearly saturate the STREAM bandwidth.
The matrix-free discretizations achieve much higher throughput because they store less data per DoF.
This model at order 2 has about \DoF{30} per element and each element has 27 quadrature points that must store 17 \code{float64} values each, resulting in about \qty{140}{B\per DoF} (including the vectors) for the matrix-free operator, and a predicted streaming peak of a bit under \GDoFs{6}.
About half of that is achieved, with the discrepancy attributable to atomic writes, vector zeroing and copies/packing related to communication and boundary conditions, and the computation to compute gradients and apply the quadrature points operations.

Note that latency is an ever-present specter, with efficiency still rising at the point when GPU memory capacity is reached.
Moreover, many applications have strict time-to-solution requirements imposed by business, policy, or human timelines, and thus \revised{it is}{it's} informative to report the time at which, say 80\% of peak efficiency is achieved.
In the subsequent section, we will place time on \revised{the}{} $x$ axis, allowing us to compare efficiency of different machines and different parallel scale.

\subsection{Nonlinear solves}
To test the efficiency of end-to-end nonlinear solves, we choose the model from \autoref{fig:schwarz-p} with (nondimensionalized) parameters thickness 0.2, Young's modulus 1, and Poisson ratio 0.3, fixed to the left wall with a compressive traction of 0.02 applied from the right.
This produces approximately 12\% strain at every resolution, which is just shy of where plastic yielding occurs for photopolymer additively manufactured products of these models \cite{maskery2018insights}.
\revised{This model requires 5 to 7}{This model requires 6 or 7} Newton iterations \revised{across the range of resolutions}{at every resolution}, with each linear solve needing 9 to 25 preconditioned CG iterations to converge to a relative tolerance of $10^{-3}$ in the natural norm, with CG condition number estimates from 9.5 to 61 (mostly less than 15 iterations and condition numbers less than 20; depending on the Newton step).

\begin{figure}
  \centering
  \includegraphics[width=.45\textwidth]{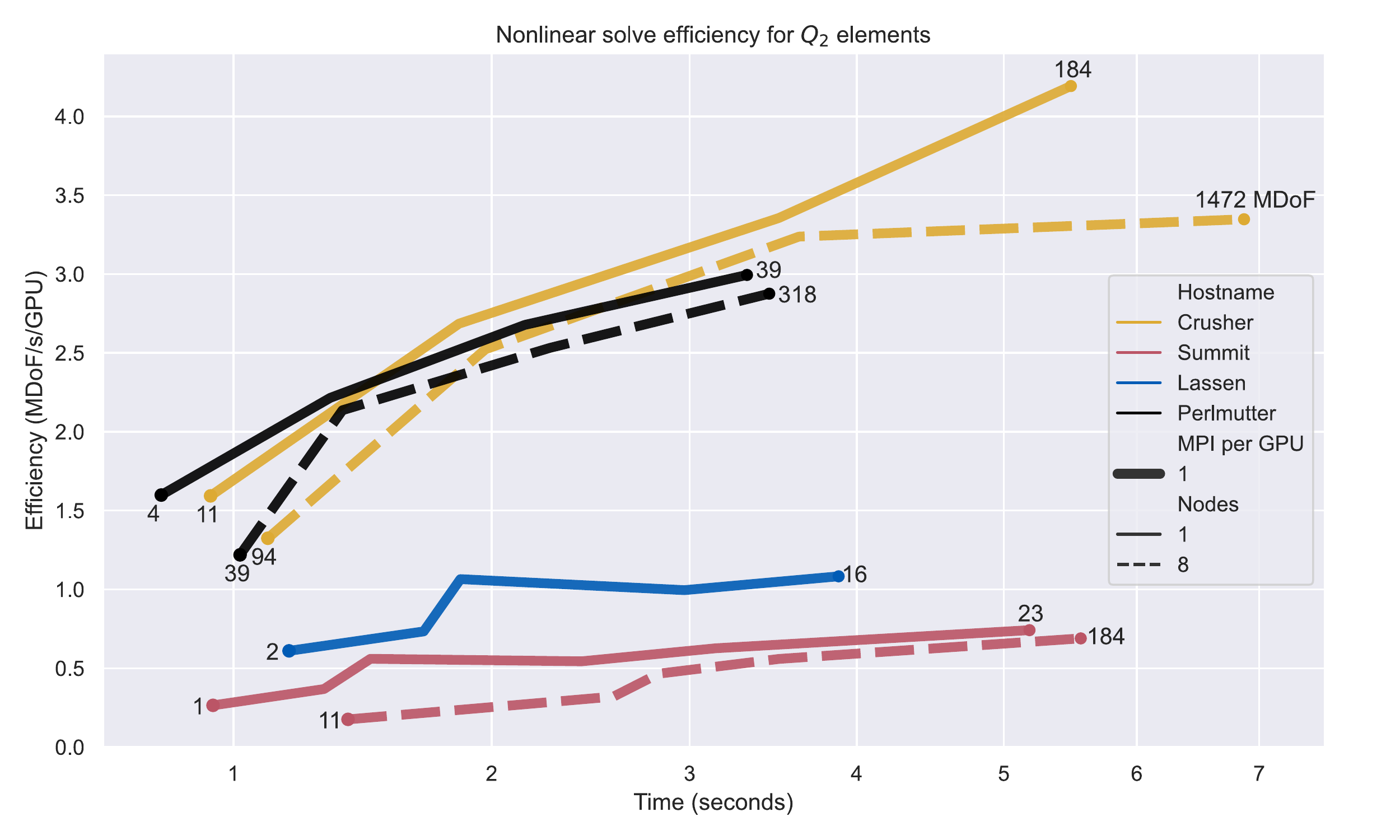}
  \caption{\revised{Efficiency per Newton iteration}{Total nonlinear solve efficiency} versus time for $Q_{2}$ finite elements using matrix-free Newton-Krylov with $p$-MG preconditioning and BoomerAMG coarse solve.
    \revised{Problem sizes (in MDoF) are annotated for the minimum and maximum sizes for each host and number of nodes combination.}{}
    The impact of latency is ever-present, with memory capacity limiting the right end of each curve.}
  \label{fig:q2-snes}
\end{figure}

\begin{figure}
  \centering
  \includegraphics[width=.45\textwidth]{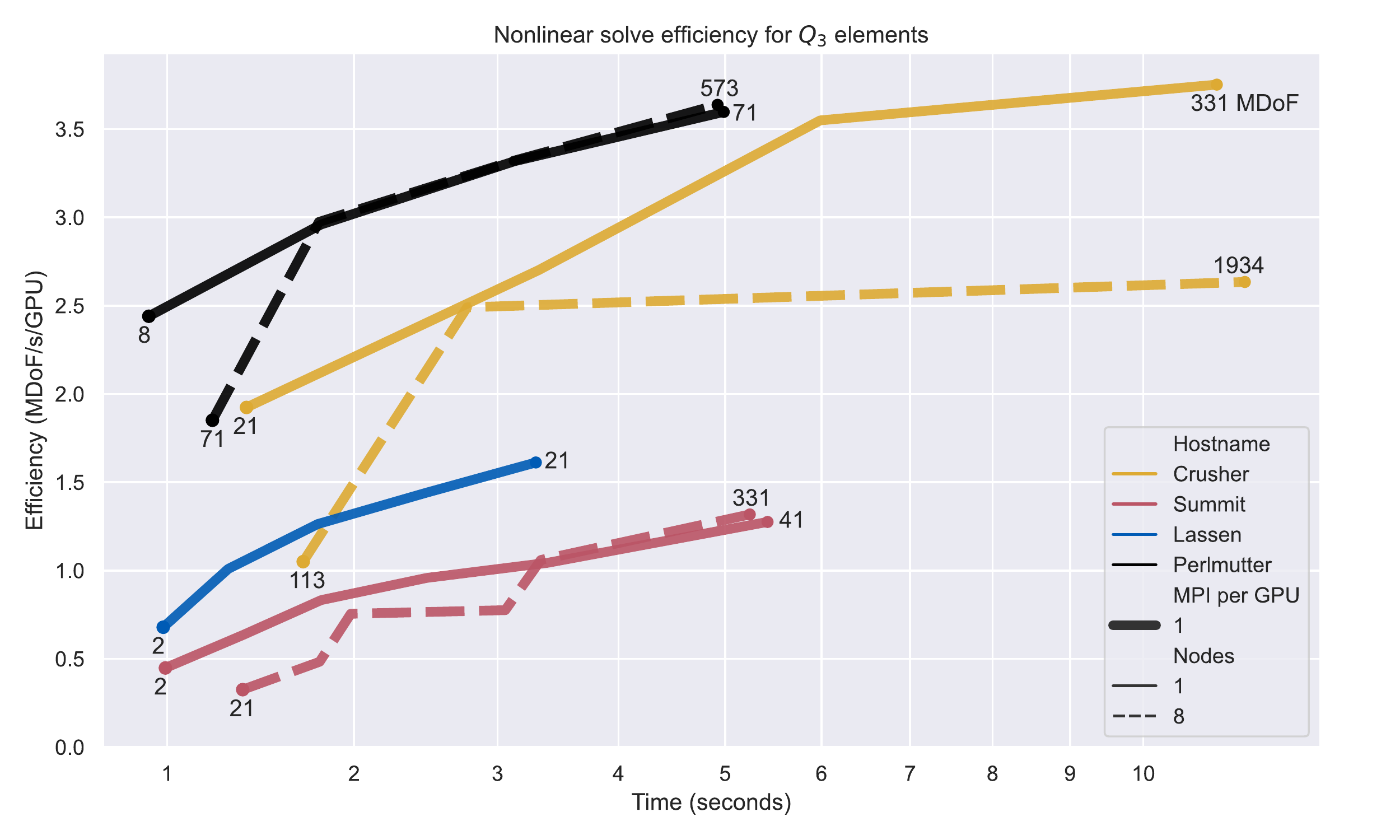}
  \caption{\revised{Efficiency per Newton iteration}{Total nonlinear solve efficiency} versus time for $Q_{3}$ finite elements using matrix-free Newton-Krylov with $p$-MG preconditioning and BoomerAMG coarse solve.
    \revised{Problem sizes (in MDoF) are annotated for the minimum and maximum sizes for each host and number of nodes combination.
    Ideal weak scaling is evident on Perlmutter for Newton step time above \qty{1.8}{\second} where the 1-node and 8-node curves coincide, while communication latency leads to degradation at the smallest problem sizes (\MDoF{2}/GPU with time around \qty{1}{\second}).}{}}
  \label{fig:q3-snes}
\end{figure}

We sweep through a range of Primitive model extents up to $20^3$ per node of Crusher (184 MDoF), solve each model, and plot efficiency versus time per Newton iteration for $Q_{2}$ and $Q_{3}$ elements in \autoref{fig:q2-snes} and \autoref{fig:q3-snes}.
The $Q_{2}$ model is as depicted in \autoref{fig:schwarz-p} with 2 refinements and 2 extruded layers, while the $Q_3$ model uses only one extruded layer to achieve somewhat better accuracy; see \autoref{table:schwarz_extrusion}.
In such plots, perfect weak scaling would have the 1-node and 8-node curves on top of each other, with strong scaling limits visible in the minimum time at which acceptable efficiency can be achieved.
This human-centric figure is meant to assist the analyst with cloud or HPC access in choosing an efficiency-versus-time tradeoff.
For example, one may look at \autoref{fig:q3-snes} and decide that \revised{under \qty{2}{\second} per Newton iteration}{} (\revised{about \qty{10}{\second}}{10 seconds} for the total nonlinear solve) delivers an acceptable efficiency\revised{-time tradeoff}{}.
Examining the Perlmutter curve with about \revised{3}{0.6} MDoF/s/GPU at \revised{2}{10} s, the target problem would be scaled to about 6 MDoF/GPU.
\revised{%
  The 1-node and 8-node Perlmutter curves lie on top of each other here, indicating that one can solve a \MDoF{24} problem on one node (4 GPUs) with the same efficiency as a \MDoF{192} problem on 8 nodes.
  Note that AMG requires a deeper V-cycle for the larger problem size, but this latency impact is hidden at the \qty{2}{\second} solve time with $Q_{3}$ elements.
  Compare with \autoref{fig:q2-snes}, in which there is a slight efficiency penalty to the weak scaling since a greater fraction of the solve time is spent in AMG when using $Q_{2}$ elements.
}{%
  For a problem size of 192 MDoF intending to keep the solve time of 10 s fixed, this calculation would result in choosing 8 nodes (4 GPUs/node) and deliver a 12 second solve time (due efficiency of about 0.5 MDoF/s/GPU for the 8-node Perlmutter line).
}
The solve can be made somewhat faster by using more GPUs (with some drop in efficiency) or more efficient by using fewer GPUs (while needing to wait longer).
Different architectures can readily be compared in this metric by normalizing the efficiency axis by energy (DoF/joule) or monetary (DoF/dollar) cost.

\begin{figure}
  \centering
  \includegraphics[width=\linewidth]{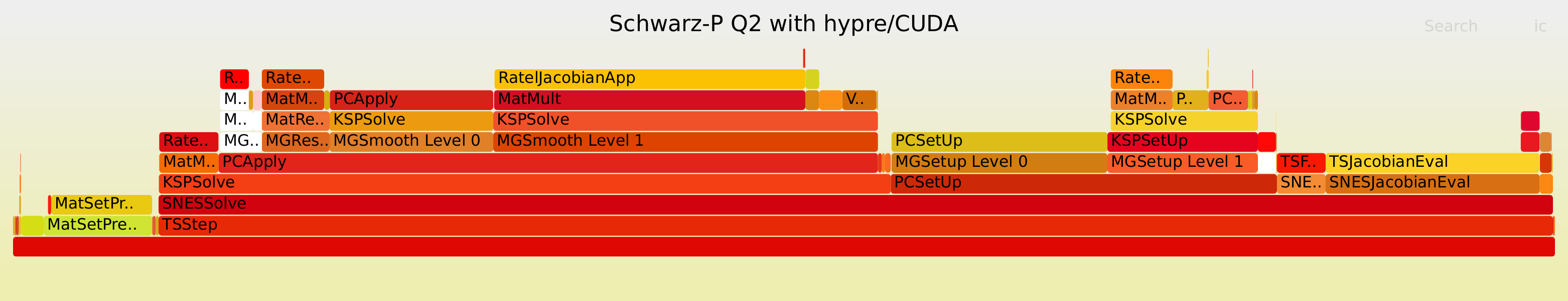}
  \caption{Flame graph for a typical setup and solve with $Q_{2}$ elements.
    \revised{The two dominant costs are preconditioner application (left half; part of the linear solve) and setup (center-right third).}{}
    The coarse solve (Level 0 and above) takes about half the time of the fine (Level 1) \revised{and coarse setup (AMG) takes more time than fine $p$-MG setup}{}, despite the fine having 8x more DoF.}
  \label{fig:q2-flame}
\end{figure}

\begin{figure}
  \centering
  \includegraphics[width=.45\textwidth]{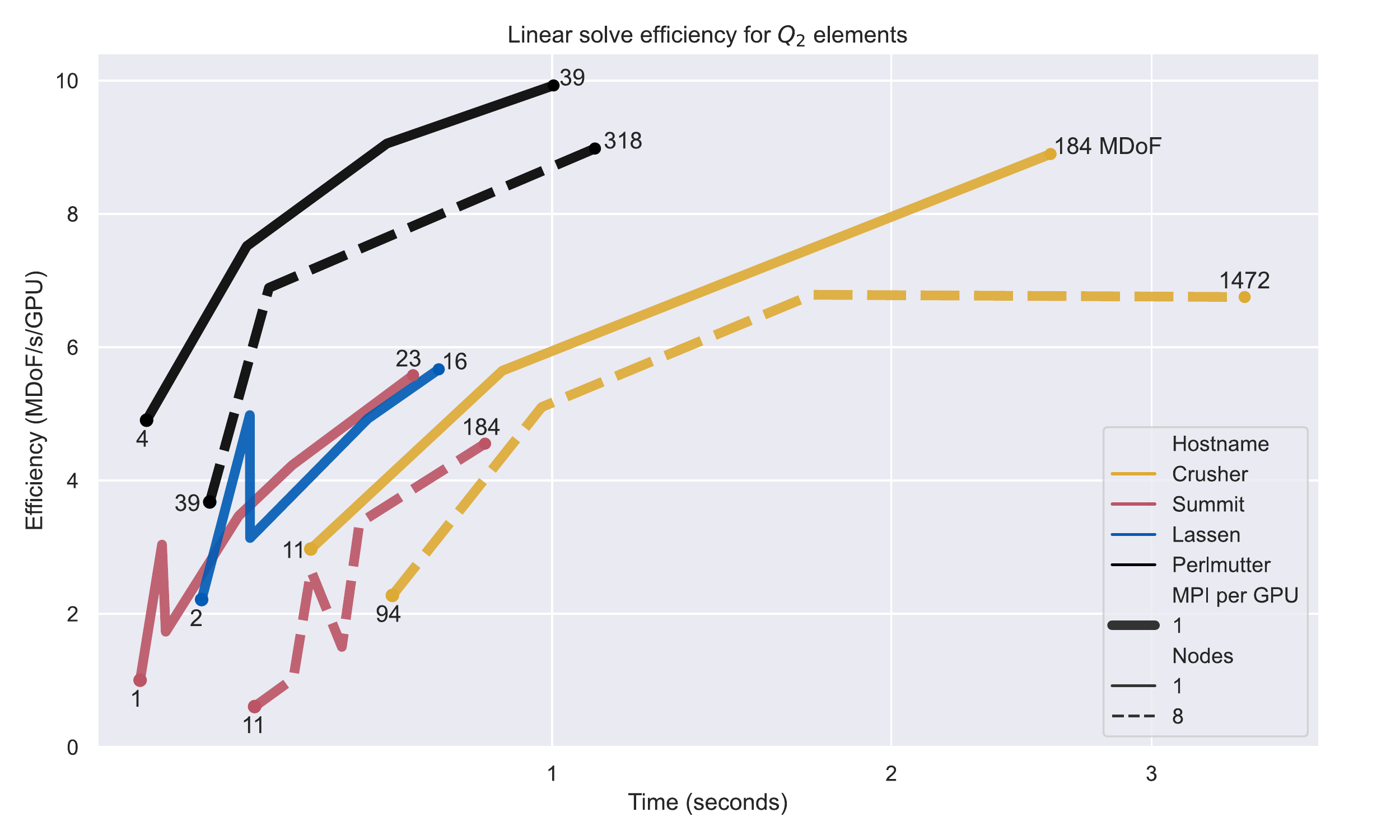}
  \caption{Linear solve efficiency spectrum for $Q_{2}$ finite elements using matrix-free Newton-Krylov with $p$-MG preconditioning and BoomerAMG coarse solve. The times and efficiencies are \revised{per Newton iteration}{summed over all Newton steps}.
  \revised{Problem sizes (in MDoF) are annotated for the minimum and maximum sizes for each host and number of nodes combination.}{}}
  \label{fig:q2-ksp}
\end{figure}

\begin{figure}
  \centering
  \includegraphics[width=.45\textwidth]{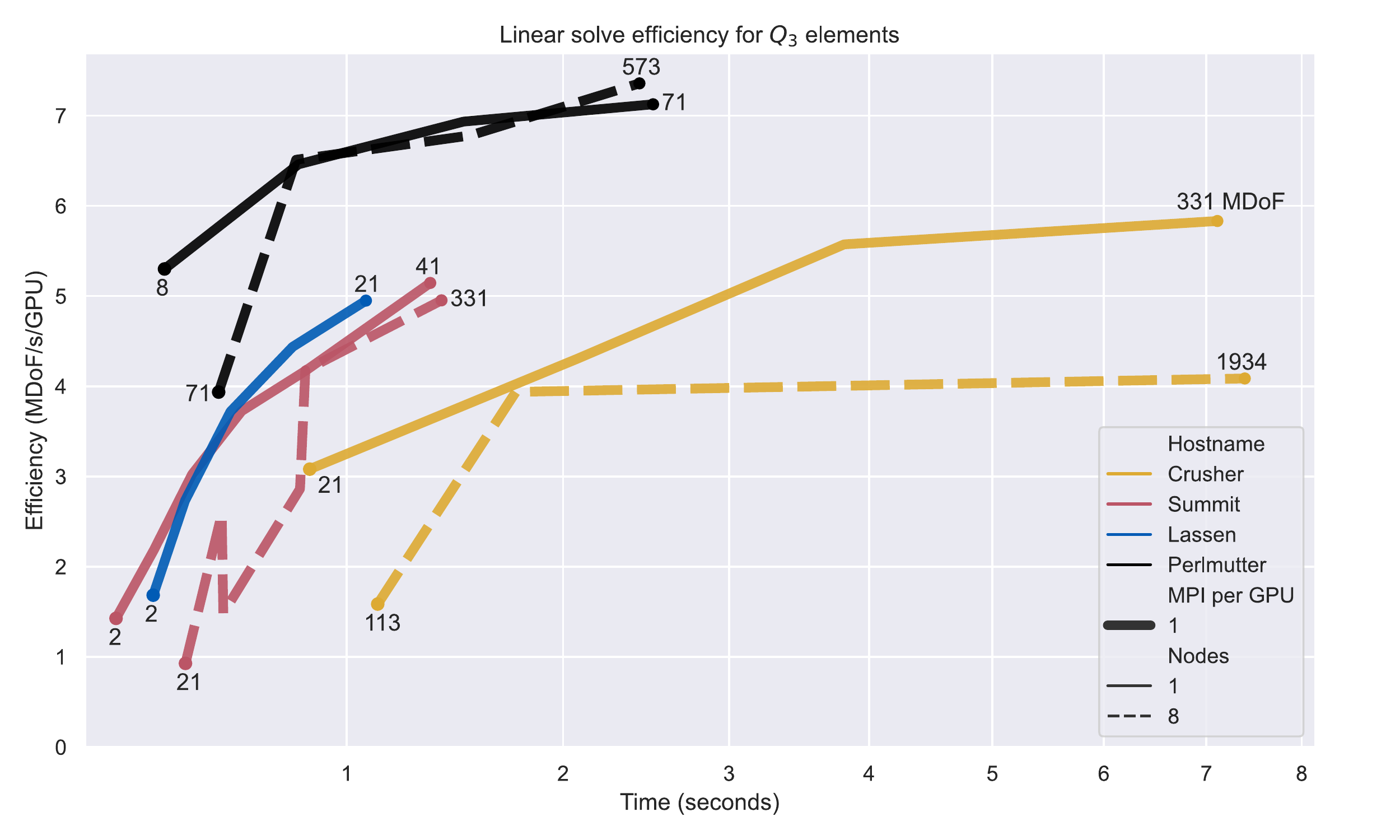}
  \caption{Linear solve efficiency spectrum for $Q_{3}$ finite elements using matrix-free Newton-Krylov with $p$-MG preconditioning and BoomerAMG coarse solve. The times and efficiencies are \revised{per Newton iteration}{summed over all Newton steps}.
  \revised{Problem sizes (in MDoF) are annotated for the minimum and maximum sizes for each host and number of nodes combination.}{}}
  \label{fig:q3-ksp}
\end{figure}

\autoref{fig:q2-flame} depicts the relative costs and relationship of the major phases, with linear solve (\code{KSPSolve}) and preconditioner setup (\code{PCSetUp}) dominating.
The linear solve (\autoref{fig:q2-ksp} and \ref{fig:q3-ksp}) is communication-intensive since each preconditioner application goes through a V-cycle (with an increasing number of levels as the model gets larger).

\begin{figure}
  \centering
  \includegraphics[width=.45\textwidth]{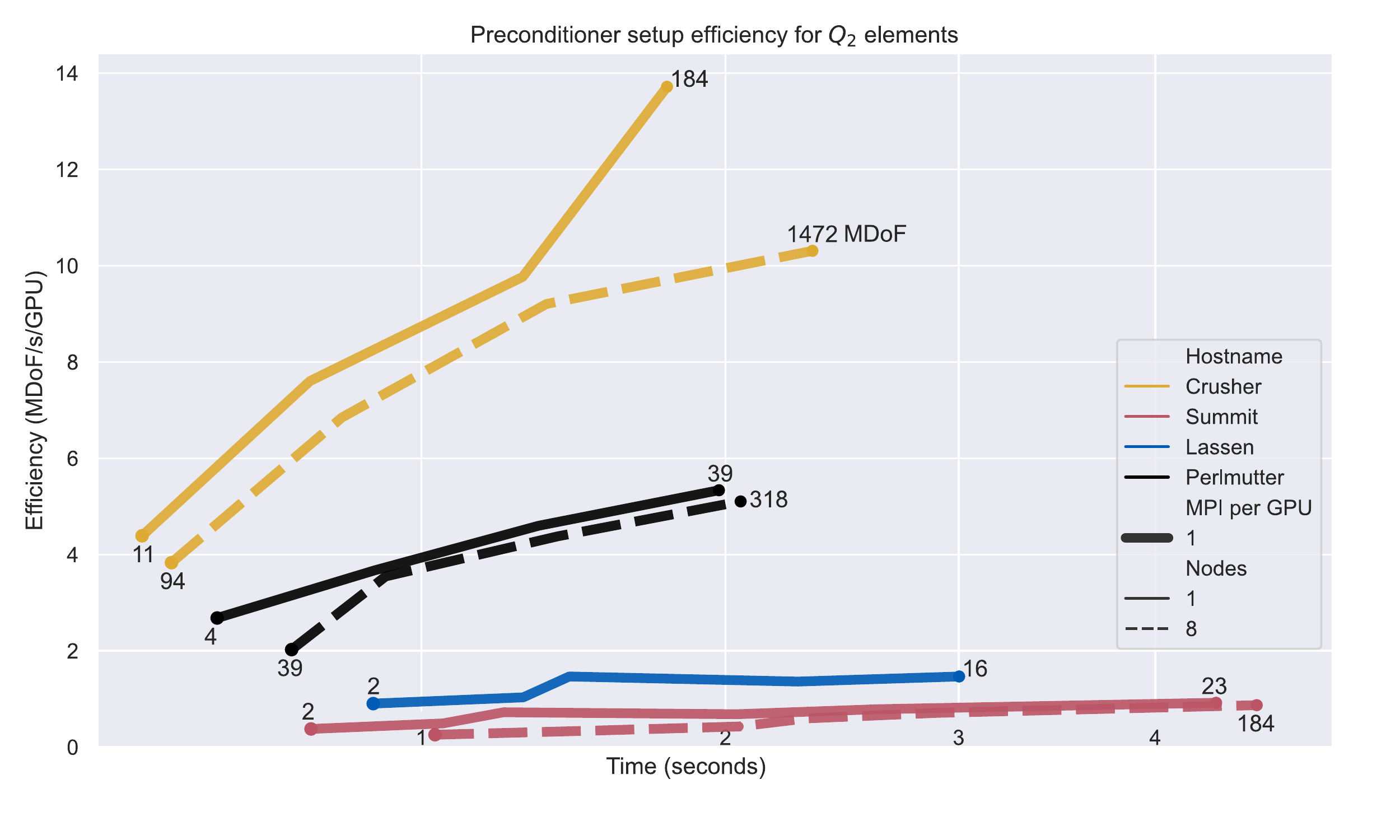}
  \caption{Preconditioner setup efficiency spectrum for $Q_{2}$ finite elements using matrix-free Newton-Krylov with $p$-MG preconditioning and BoomerAMG coarse solve. The times and efficiencies are \revised{per Newton iteration}{summed over all Newton steps}.
  \revised{Problem sizes (in MDoF) are annotated for the minimum and maximum sizes for each host and number of nodes combination.}{}}
  \label{fig:q2-pcsetup}
\end{figure}

\autoref{fig:q2-pcsetup} considers preconditioner setup, which consists of algebraic multigrid analysis and Galerkin products as well as a few Krylov iterations to calibrate the smoothers.
Assembly of the coarse Jacobian (\code{SNESJacobianEval}\revised{; far-right yellow}{} in \autoref{fig:q2-flame}) exhibits nearly perfect problem-size independence at about \revised{20--25\,MDoF/s/GPU}{4 GDoF/s/GPU} (\revised{}{over all solves,} with $Q_{2}$ elements) on Crusher and Perlmutter, and is thus always less than 20\% overhead.
The relative cost of Jacobian assembly and preconditioner setup both decrease when going to $Q_{3}$ elements because the coarse problem is a smaller fraction of the fine problem size.
When solving transient problems or nonlinear problems by quasi-Newton methods, the preconditioner setup can often be reused across many solves and thus this phase becomes \revised{insignificant in comparison to the increasingly}{} dominant \revised{linear solve}{}.
Although these figures show Newton-Krylov methods due to their familiar properties and diagnostics, we observe up to 2x performance improvement when using L-BFGS as described in \autoref{sec:iterative}, and recommend testing it on representative problems.

\revised{%
  In general, we observe greater volatility in the ``strong scaling'' regime at the left edge of Figures \ref{fig:q2-snes} to \ref{fig:q2-pcsetup}.
  Most configurations reach high efficiency weak scaling (solid and dotted lines very close) as the problem size per GPU increases, leading to Newton solve times increasing to around \qty{2}{\second} and higher.
  This efficient weak scaling is usually realized at smaller (faster) solves than where performance plateaus, indicating that 1-node architectural latencies are a more insidious performance obstacle than multi-node communication.
  Although Crusher exhibits a regime of efficient scaling, the efficiency degrades at the largest problem sizes.
  This effect is not present on other machines and our profiling points to network degradation not identifiable in microbenchmarks (or smaller problem sizes) that we hope will be resolved in the MPI implementation or tuning.
  Specifically, the majority of the degradation is attributable to point-to-point messaging and Jacobian assembly communication that performance models indicate should be cheap relative to volume work because these huge subdomains have low surface area to volume ratio.
}{}

\subsection{Robustness}
We now explore under what circumstances the $p$-MG limits convergence versus when \revised{it is limited by}{it's} the AMG coarse solve.
In order to make direct solvers affordable, we consider the Primitive model with extent (8,2,2) under 0.001 tension and report the iteration count and condition number from the first solve using Hypre, GAMG, and Cholesky.
\autoref{tab:robustness} investigates convergence for the mildly stretched elements in the original thickness 0.2 and more stretched in thickness 0.05, both with two layers.
We have fixed solver parameters to be representative of problems with both well-shaped and stretched elements; better convergence on stretched models can be obtained by tuning threshold and smoothing parameters for the worst quality elements, at the expense of degraded convergence for better-shaped elements (left column).

\begin{table}
  \centering
  \caption{Preconditioner robustness for stretching experiment.}
  \label{tab:robustness}
  \begin{tabular}{llrrrr}
    \toprule
    & & \multicolumn{2}{c}{Thickness 0.2} & \multicolumn{2}{c}{Thickness 0.05} \\
    Order & Preconditioner & Its & Cond & Its & Cond \\
    \midrule
    1 & Hypre & 12 & 10 & 27 & 91 \\
    1 & GAMG & 19 & 26 & 111 & 942 \\
    2 & Hypre & 14 & 14 & 55 & 253 \\
    2 & GAMG & 32 & 72 & 293 & 6545 \\
    2 & p-MG, Hypre & 19 & 59 & 74 & 475 \\
    2 & p-MG, GAMG & 12 & 10 & 105 & 854 \\
    2 & p-MG, Cholesky & 8 & 7 & 52 & 436 \\
    3 & p-MG, Hypre & 15 & 18 & 62 & 386 \\
    3 & p-MG, GAMG & 12 & 11 & 104 & 818 \\
    3 & p-MG, Cholesky & 9 & 8 & 48 & 369 \\
    \bottomrule
  \end{tabular}
\end{table}


\subsection{Usability via Automatic Differentiation}\label{libceed-enzyme}
Efficient use of matrix-free methods requires quadrature-point based linearization (``partial assembly'') of  forward (and possibly adjoint) operators. 
While many problems have structure \cite{brown2010efficient,davydov2020matrix} that can reduce the memory footprint and operation count, it can be tedious to find these formulations and \revised{it is}{it's} onerous to have to develop the nonlinear residual and Jacobian action synchronously.
Automatic differentiation (AD) tools simplify this process, automating the Jacobian action so that only the nonlinear forward model needs to be written by a human.
Enzyme \cite{moses2021enzymegpu} is a new LLVM plugin with GPU support that provides split forward and reverse mode AD on LLVM intermediate representation (IR).

We investigate applicability and performance computing the Jacobian action using Enzyme's new (version 0.0.29) split forward-mode capability to provide derivatives of Neo-Hookean models.
Split mode populates a ``tape'', which contains opaque intermediate values at quadrature points, and is stored in ordinary libCEED arrays (output from residual computation and input to Jacobian evaluation).
The material model expressed in current configuration \eqref{Kirchhoff-stress} is too simple for this test so we include tests of the same model expressed in initial configuration: the second Piola-Kirchhoff stress as a function of the Green-Lagrange strain, $\bm S(\bm E)$.
This model contains matrix inverses and thus its analytic derivative uses the identity $\diff \bm C^{-1} = - \bm C^{-1}\diff \bm C\, \bm C^{-1}$, but this is not known to Enyzme.
Enzyme identifies a straightforward and relatively low-memory representation (small tape) given the provided structure, and the resulting vectorized code is on par with naive hand-written code that doesn't exploit symmetries and cancellation.
\autoref{table:different-storage} compares total solve time (over many steps) on a small cube mesh with $3630$ DoF (fits in cache, thus stresses flops) on a single process of an AMD EPYC 7452.
The initial configuration cases need to store initial configuration geometry $\nabla_{X}\bm\xi$ and quadrature-weighted determinants $W$, while the current configuration maps directly to the solution-dependent current configuration $\nabla_{x}\bm\xi$.
Since Enzyme is language-agnostic (by virtue of operating on LLVM IR), this opens the door to constitutive modeling in safer/more convenient languages, such as Rust and Julia, with no impact on execution performance or environment.

\begin{table}
\centering
\caption{Performance for different Jacobian representations in Neo-Hookean hyperelasticity.
Stored values before the semicolon are constant data while those after are a byproduct of residual evaluation.}
  \label{table:different-storage}
  \begin{tabular}{cccc}
    \toprule
    Problem & Storage & Scalars & Time (s) \\
    \midrule
    \code{current} & $W; \nabla_x \bm\xi, \bm \tau, \lambda \log J$ & 17 & 7.097 \\
    \code{initial native} & $\nabla_X\bm\xi, W; \nabla_X \bm u$ & 19 & 11.556 \\
    \code{initial tuned} & $\nabla_X\bm\xi, W; \nabla_X \bm u, \bm C^{-1}, \lambda \log J$ & 26 & 9.498 \\
    \code{initial AD} & $\nabla_{X}\bm\xi, W; \nabla_X \bm u, S$, \code{tape} & 31 & 10.661 \\
    \bottomrule
  \end{tabular}
\end{table}


\section{Discussion}\label{sec:discussion}

High order methods have thus far made little impact on industrial practice of structural engineering primarily due to performance consequences of traditional sparse matrix abstractions, which offer increasingly poor utilization of modern hardware.
We have shown that this performance landscape is inverted by changing data structures to matrix-free representations with linearization defined at quadrature points, for which high order methods are significantly cheaper per DoF.
\revised{The 1-node Crusher examples solve problems of similar size (hundreds of MDoF) and complexity to the implicit structural mechanics problems in the 2002 \cite{bhardwaj2002salinas} and 2004 \cite{adams2004ultrascalable} Gordon Bell Prizes, which are still considered large in industrial and research structural mechanics practice.
  Our methods enable pragmatic use of $Q_{2}$ and $Q_{3}$ elements while delivering much faster time to solution.
  While we have focused our study here on problems of size less then \GDoF{2} to maximize interpretability and relevance to practitioners, the algorithms are scalable to much larger problems and node counts.}{}

Similar structure has previously been exploited by \cite{arbenz2008scalable} to reduce memory requirements by not storing the fine-grid matrix for bone structure analysis, while still constructing prolongation operators for smoothed aggregation AMG preconditioning.
This work was restricted to linear elasticity on voxelized meshes and the setup and solve time was somewhat longer than with standard assembled methods.
Other recent work \cite{davydov2020matrix} \revised{(based on the fast matrix-free operators \cite{kronbichler2012generic} in the deal.II library)}{} used geometric $h$-multigrid for high order finite elements applied to hyperelasticity on CPUs, showing excellent performance for a matrix-free methods relative to assembled methods.
In particular, the matrix-free iteration counts were found to be much smaller than AMG applied directly to the assembled high-order discretization, and each iteration was cheaper by virtue of the matrix-free data structures.
In that work, the high order discretization was preserved on nested coarse grids, which limits applicability to problems with high geometric complexity.
Non-nested geometric multigrid \cite{feng1997nonnested,adams2002evaluation,kong2016multilevel} could be extended to high order elements, but studies of such methods in complex geometry \cite{adams2002evaluation} have encountered robustness problems relative to algebraic multigrid.

The matrix-free $p$-multigrid approach presented here offers the robustness of low-order algebraic multigrid with much higher efficiency per DoF and simulation time to reach engineering tolerances.
The use of high order elements has the additional benefit that coarser meshes can be used, thus reducing preprocessing time and I/O costs related to element topology, though it requires more attention to element quality at the mesh generation stage.
We find that quadratic geometry is often sufficient for large deformation with quadratic and cubic solution spaces, and thus these methods can be used with existing meshing and visualization tools, though tailoring to $p$-version finite element efficiency \cite{luo2004automatic} is beneficial.
When cubic and higher order meshes are needed, one can use Gmsh \cite{geuzaine2009gmsh} to generate arbitrary order meshes, but many popular mesh formats support at most second order elements and there is a need for improving data representation standards and postprocessing/visualization tools to better support high order geometry and solution fields \cite{remacle2007efficient,glvis-tool,rasquin2021post}.

We find that the algorithms here provide substantial benefit already for quadratic elements, and thus is a viable drop-in technique any time quadratic geometry can be used, and sometimes even for linear geometry elements.
One can switch from linear to quadratic elements on the same mesh for about double the cost, despite 8 times more DoFs.
The method is applicable for almost any problem in which the coarsest geometry-resolving mesh is not accurate enough for simulation with linear elements.
Despite equivalent asymptotic convergence in the presence of singularities for high order methods, the combination of constants for approximation and algorithmic implementation efficiency often leads to an order of magnitude reduction in cost to reach engineering tolerances, offering a transformative opportunity to make batch simulations interactive and greatly expand the use and fidelity of solid mechanics simulation in science and industry.
\revised{While the methods require revisiting the traditional centrality of sparse matrices for implicit finite element analysis, most of the algorithmic structure remains intact (with a new economy that inverts key instances of conventional wisdom to enable further efficiency gains).
Note that libCEED was designed for use in legacy software; adoption by conventional CPU-based implicit FEA software is mostly a matter of calling material models in libCEED Q-functions and modest data structure abstraction in the solver via standard interfaces provided by PETSc and similar libraries.}{}

One limitation to the matrix-free $p$-multigrid technique is that Chebyshev/Jacobi smoothing degrades for highly stretched elements, such as appear in volumetric discretizations of shell structures.
One needs either semi-coarsening or block/line smoothers to make multigrid convergence uniform on such models, neither of which is especially convenient in the present framework.
We note that shell structures usually have small vertex separators and thus direct solvers and parallel adaptive BDDC solvers such as PETSc's PCBDDC \cite{zampini2016pcbddc} offer sharp convergence guarantees at manageable cost.

\section*{Acknowledgment}
This research is supported by the Exascale Computing Project (17-SC-20-SC), a collaborative effort of two U.S. Department of Energy organizations (Office of Science and the National Nuclear Security Administration) responsible for the planning and preparation of a capable exascale ecosystem, including software, applications, hardware, advanced system engineering and early testbed platforms, in support of the nation's exascale computing imperative.
This research is supported by the U.S. Department of Energy, Office of Science, Office of Advanced Scientific Computing Research under contract DE-AC02-06CH11357 and Award Number DE-SC0016140.
The authors acknowledge support by the Department of Energy, National Nuclear Security Administration, Predictive Science Academic Alliance Program (PSAAP) under Award Number DE-NA0003962.
This research used resources of the Oak Ridge Leadership Computing Facility at the Oak Ridge National Laboratory, which is supported by the Office of Science of the U.S. Department of Energy under Contract No. DE-AC05-00OR22725.
This research used resources of the Livermore Computing Facility at the Lawrence Livermore National Laboratory.
This research used resources of the National Energy Research Scientific Computing Center, which is supported by the Office of Science of the U.S. Department of Energy under Contract No. DE-AC02-05CH11231.
We thank PETSc, hypre, and libCEED developers, especially Mark Adams, Barry Smith, Stefano Zampini, Victor Paludetto Magri, Veselin Dobrev, Yohann Dudouit, and Tzanio Kolev, for collaboration on software development and algorithmic tuning.
This research used Paraview, Seaborn, Matplotlib, and Pandas for data analysis and visualization.

\bibliographystyle{unsrt}
\bibliography{perf-portable-solids_refs.bib}

\end{document}